\documentclass[amssymb,aps,prd,floatfix,10pt,superscriptaddress,nofootinbib]{revtex4}
\usepackage{graphicx}
\usepackage{float}
\usepackage{mathtools}
\usepackage{mathrsfs}
\usepackage{tkz-euclide}
\usepackage{siunitx}
\usepackage{multirow}
\usepackage[caption=false]{subfig}
\usepackage[x11names]{xcolor}
\usepackage[colorlinks=true,linkcolor=blue,citecolor=Blue4,urlcolor=Blue4]{hyperref}
\usepackage{orcidlink}

\begin{document}
\raggedbottom

\title{Kerr geodesics on conformal projection diagrams}

\author{Przemys{\l}aw Podle\'{s}ny \protect\orcidlink{0009-0002-9222-1101}}
\email{przemek.podlesny@doctoral.uj.edu.pl}
\affiliation{Szko{\l}a Doktorska Nauk \'{S}cis{\l}ych i~Przyrodniczych, Uniwersytet Jagiello\'{n}ski, \L ojasiewicza 11, 30-348 Krak\'ow, Poland}
\affiliation{Instytut Fizyki Teoretycznej, Uniwersytet Jagiello\'{n}ski, {\L}ojasiewicza 11, 30-348 Krak\'ow, Poland}

\author{Patryk Mach \protect\orcidlink{0000-0003-1493-8668}}
\email{patryk.mach@uj.edu.pl}
\affiliation{Instytut Fizyki Teoretycznej, Uniwersytet Jagiello\'{n}ski, {\L}ojasiewicza 11, 30-348 Krak\'ow, Poland}

\begin{abstract}
We provide a rigorous representation of timelike geodesics in the maximally extended Kerr spacetime, using projection diagrams introduced by Chru\'{s}ciel, \"{O}lz, and Szybka. In contrast to standard conformal diagrams, projection diagrams allow for plotting general causal curves (not necessarily confined to the two-dimensional surface associated with the diagram), preserving their causal characteristics. Our analysis utilizes two systems of horizon-penetrating coordinates (both ingoing and outgoing) to ensure a regular description of geodesics crossing the event and Cauchy horizons. On the other hand, at most four causal blocks can be joined smoothly by a Kruskal-Szekeres-type construction proposed for projection diagrams. Consequently, depicting a trajectory spanning several causal blocks may require more than one diagram, as the conformal factor either vanishes or diverges at the block boundaries.
\end{abstract}

\maketitle

\section{Introduction}
Conformal diagrams, pioneered by Penrose \cite{RP1963,RP1964} and further developed by Carter \cite{BC1966}, have revolutionized the ways in which the causal and global structures of spacetimes are visualized. On the other hand, they are two-dimensional by construction. Consequently, when a physical process is presented on a conformal diagram, a large portion of the relevant information is often discarded. Although this is not necessarily problematic for highly symmetric spacetimes (e.g., in spherical symmetry), visualization of processes occurring in less symmetric ones remains challenging. In this regard, the Kerr spacetime can serve as a standard example. Conformal diagrams for the Kerr spacetime are usually drawn for points at the symmetry axis \cite{HE} but rarely for other choices (e.g., the equator). To be precise, conformal diagrams are constructed for two-dimensional Lorentzian manifolds $\mathcal{N}$---hypersurfaces in the original spacetime $(\mathcal{M}, g)$. In the case of the Kerr spacetime, they can be selected by fixing the polar and azimuthal angle, in a given coordinate system. Unfortunately, conformal diagrams constructed for different hypersurfaces $\mathcal{N}$ can differ qualitatively.

Our work has been motivated by the seemingly simple task of illustrating timelike geodesics on a conformal diagram of the maximally extended Kerr spacetime. This poses an additional difficulty. A diagram constructed from a hypersurface $\mathcal{N}$ can be compatible with trajectories confined to this hypersurface, but hardly for general trajectories. This situation was partially remedied by the projection diagrams, introduced by Chru\'{s}ciel, \"{O}lz, and Szybka in \cite{COS2012}. In what follows, we will refer to this construction as the COS projection diagrams. Roughly speaking, a COS projection diagram constitutes a map from (a subset of) the spacetime $(\mathcal M, g)$ to the two-dimensional Minkowski spacetime such that the following two conditions hold: i) Every smooth timelike curve on the diagram is a projection of a smooth timelike curve in the spacetime $(\mathcal M, g)$. ii) Every smooth timelike curve in the spacetime $(\mathcal M, g)$ is projected on a smooth timelike curve in the diagram. Additional technical details in the definition of COS projection diagrams (introduced in \cite{COS2012} and repeated in Sec.\ \ref{sec:projdiag}) allow one to exclude unwanted regions from the spacetime $(\mathcal M, g)$. For the Kerr spacetime, which is of our interest in this work, this applies to the so-called Carter Time Machine (CTM) region containing closed timelike curves.

In this work, we illustrate geodesic motion in the Kerr spacetime on the COS projection diagrams. Even within this framework, this task leads to two unwanted difficulties. The first one is related to the construction of COS projection diagrams, which is based on the standard Kruskal-Szekeres-type method \cite{Kruskal1960,Szekeres,W1970}, which in turn generally allows for a smooth gluing of four causal blocks (diamonds) only. Depicting a trajectory spanning across a larger number of blocks requires separate gluings, realized by different choices of the conformal factor. The second difficulty is related to the behavior of geodesics that cross Kerr horizons. In our approach, they are described in terms of horizon-penetrating Kerr coordinates. This requires two separate coordinate systems, allowing for a regular representation of ingoing and outgoing geodesics, respectively. Consequently, a depiction of a single geodesic incoming from infinity, plunging into the black hole, encountering a turning point beneath the inner horizon, and escaping the ``white hole'' in subsequent causal blocks may require three different diagrams and two coordinate systems to describe the geodesic itself. We emphasize that we aim for a quantitative, numerical representation rather than a qualitative schematic drawing. 

This work is organized as follows. In Section \ref{sec:baseconv} we review basic facts related to Kerr geometry and provide a brief characterization of the CTM set. Section \ref{sec:projdiag} outlines the construction of COS diagrams for the Kerr spacetime. This is essentially the material known from \cite{COS2012}, to which we add some technical aspects needed in our work. In Section \ref{sec:geoeq} we provide a description of causal Kerr geodesics in horizon-penetrating coordinates. The solutions of the resulting equations are discussed in Sec.\ \ref{sec:geosol}. This part is mainly based on \cite{B2025}. Compared to the results obtained in \cite{B2025}, we add a description of causal Kerr geodesics in the horizon-penetrating outgoing coordinates. In Section \ref{sec:examples} we provide examples of timelike geodesics plotted on COS projection diagrams. These results were obtained numerically, using \textit{Wolfram Mathematica} \cite{Wolfram}. A short summary can be found in Sec.\ \ref{sec:conclusions}.

We assume the signature of the metric $(-,+,+,+)$ and use geometric units with $c=G=1$, where $c$ denotes the speed of light and $G$ is the gravitational constant.

\section{Basic conventions, Carter time-machine set}
\label{sec:baseconv}

\subsection{The Kerr metric}

The Kerr metric, expressed in Boyer-Lindquist coordinates $(t,r,\theta,\phi)$, has the form \cite{BL1967}
\begin{align}
\begin{aligned}
g=&-\left(1-\dfrac{2Mr}{\Sigma}\right)dt^{2}+\dfrac{\Sigma}{\Delta}dr^{2}+\Sigma d\theta^{2}-\dfrac{4Mar}{\Sigma}\sin^{2}{\theta}dtd\phi \\
&+\left(r^{2}+a^{2}+\dfrac{2Ma^{2}r\sin^{2}{\theta}}{\Sigma}\right)\sin^{2}{\theta}d\phi^{2},
\end{aligned}
\label{KerrBL}
\end{align}
where
\begin{subequations}
    \begin{align}
\Sigma&=r^{2}+a^{2}\cos^{2}{\theta}, \\
\Delta&=r^{2}-2Mr+a^{2}.
\end{align}
\end{subequations}
Here, $a=J/M$ denotes the Kerr spin parameter, with $J$ and $M$ representing the angular momentum and the mass of the black hole, respectively. The inner and outer horizons are located at the roots of the equation $\Delta=0$, given by
\begin{align}
r_{\pm}=M\pm\sqrt{M^{2}-a^{2}}.
\end{align}

In what follows, we shall restrict our considerations to the case where $M > 0$ and $M>\left|a\right|$. The Kerr metric admits two Killing vectors $\partial_{t}$ and $\partial_{\phi}$, reflecting the stationarity and the axial symmetry of the Kerr spacetime.

Two systems of coordinates regular at the horizons can be obtained by the following transformations of the time and azimuthal angle $(t,\phi) \to (\tilde{t}, \tilde{\phi})$:
\begin{subequations}
\label{horizonpenetrating}
\begin{align}
d\tilde{t}&=dt+\epsilon_{\mu}\dfrac{2Mr}{\Delta}dr, \\
d\tilde{\phi}&=d\phi+\epsilon_{\mu}\dfrac{a}{\Delta}dr.
\end{align}
\end{subequations}
Here, $\epsilon_{\mu}=+1$ corresponds to the ingoing coordinate system, while the opposite sign describes the outgoing one \cite{TP1974}. Alternatively, one can write
\begin{subequations}
    \begin{align}
        \tilde{t}&=t+\epsilon_{\mu}\left(r_{\ast}-r\right) + C_{t}, \label{ttilde} \\
        \tilde{\phi}&=\phi+\dfrac{\epsilon_{\mu}a}{2\sqrt{M^{2}-a^{2}}}\log{\left|\dfrac{r-r_{+}}{r-r_{-}}\right|} + C_{\phi},
    \end{align}
\end{subequations}
where
\begin{align}
    r_{\ast}(r) = r+\dfrac{2Mr_{+}}{r_{+}-r_{-}}\log{\left|\dfrac{r-r_{+}}{r_{+}}\right|}-\dfrac{2Mr_{-}}{r_{+}-r_{-}}\log{\left|\dfrac{r-r_{-}}{r_{-}}\right|}
\end{align}
satisfies $r_\ast^\prime(r) =  (r^2 + a^2)/\Delta$, except for $r = r_\pm$.

We emphasize that the integration constants $C_{t}$ and $C_{\phi}$ can be chosen separately in each of the regions $r < r_-$, $r_- < r < r_+$, and $r_+ < r$. In practice, we set $C_{t} = C_{\phi} = 0$ in this paper. We will refer to coordinates $(\tilde t, r, \theta, \tilde \phi)$ as Kerr horizon-penetrating coordinates. They allow us to express the Kerr metric as
\begin{align}
    \begin{aligned}
        g = & -\left(1-\frac{2Mr}{\Sigma}\right){d \tilde t}^2-\frac{4Mr}{\Sigma}a\sin^2\theta d \tilde t d \tilde \phi + \epsilon_{\mu}\frac{4Mr}{\Sigma} d \tilde t d r - 2\epsilon_{\mu} a \left(1 + \frac{2Mr}{\Sigma}\right)\sin^2\theta dr d \tilde \phi \\
        &+ \left(1+\frac{2Mr}{\Sigma}\right) dr^2 + \Sigma d\theta^2 + \frac{\left[ (r^2+a^2)^2-a^2 \Delta \sin^2\theta \right]\sin^2\theta}{\Sigma} {d \tilde \phi}^2,    
    \end{aligned}
    \label{gKerrSchild}
\end{align}
or alternatively
\begin{align}
g = -d\tilde{t}^{2} +dr^{2} -2\epsilon_{\mu}a\sin^{2}{\theta} dr d\tilde{\phi} 
 + \left( r^{2} + a^{2} \right)\sin^{2}{\theta} d\tilde{\phi}^{2}
 + \Sigma d\theta^{2} 
 + \dfrac{2M r}{\Sigma}  \left(d \tilde{t} + \epsilon_{\mu}dr - a\sin^{2}{\theta} d\tilde{\phi} \right)^{2},
\label{Eq:Kerr}
\end{align}
in the so-called Kerr-Schild form.

The vector normal to the hypersurfaces of constant time $\tilde t$ has the form
\begin{equation}
    N = \left( 1 + \frac{2 M r}{\Sigma} \right) \partial_{\tilde t} - \epsilon_\mu \frac{2 M r}{\Sigma} \partial_r,
\end{equation}
so that $N_\mu = (-1,0,0,0)$. Since $g(N,N) = - d \tilde t (N) = - \left( 1 + 2 M r /\Sigma \right)$, the vector $N$ is timelike for $r^2 + 2 M r + a^2 \cos^2 \theta > 0$, i.e., for $r \in \left(-\infty,-M-\sqrt{M^2 - a^2 \cos^2 \theta} \right) \cup \left( -M + \sqrt{M^2 - a^2 \cos^2 \theta},+\infty \right)$. In particular, it is timelike for $r > 0$. In this region, $N$ can be used to define the time orientation. We assume that it is future-directed. In the calculations presented in this paper, we test the time orientation using an alternative set of vectors:
\begin{equation}
\label{XY}
    X \coloneqq (r^2 + a^2) \partial_{\tilde t} + a \partial_{\tilde \phi}, \qquad Y \coloneqq 2 M r \partial_{\tilde t} + \epsilon_\mu \Delta \partial_r + a \partial_{\tilde \phi}. 
\end{equation}
The vector $X$ satisfies $g(X,X) = - \Sigma \Delta$. Consequently, it is timelike if $\Delta > 0$, i.e., for $r \in (-\infty,r_-) \cup (r_+,\infty)$. In contrast, $g(Y,Y) = \Sigma \Delta$, and thus $Y$ is timelike for $r \in (r_-,r_+)$. Furthermore, $g(N,X) = -d \tilde t(X) = - (r^2 + a^2) \le 0$ and $g(N,Y) = - d \tilde t(Y) = - 2 M r$, so that $g(N,Y) < 0$ for $r > 0$. Thus, both $X$ and $Y$ are future-directed in the respective regions, in which they are timelike, and $N$ also remains timelike. We extend this definition and treat the vector $X$ as future-directed also for $r \in \left( -M-\sqrt{M^2 - a^2 \cos^2 \theta}, -M + \sqrt{M^2 - a^2 \cos^2 \theta} \right)$.

In the following sections, we will mostly use the Boyer-Lindquist coordinates. Horizon-penetrating coordinates will be used to solve geodesic equations in Sec.\ \ref{sec:geoeq}.

\subsection{Carter time-machine region}

The maximal analytic extension of the Kerr metric is known to contain the so-called Carter time-machine (CTM) region, in which the Killing vector $\partial_\phi$ becomes timelike. As the orbits of $\partial_\phi$ contained in the CTM region are $2\pi$-periodic in $\phi$, they form closed timelike curves \cite{C1968}. 

Since the existence of the CTM region plays a role in the forthcoming discussion, we will now provide its short characterization. We first check the conditions under which the Killing vector $\partial_\phi$ is spacelike. Since
\begin{align}
\label{gff}
    g_{\phi \phi} = \left(r^{2}+a^{2}+\dfrac{2Ma^{2}r\sin^{2}{\theta}}{\Sigma}\right)\sin^{2}{\theta},
\end{align}
one concludes that for $r > 0$, $g_{\phi \phi} \ge 0$ and the vector $\partial_{\phi}$ is spacelike. For $r \leq 0$, we follow \cite{COS2012} and define the CTM set as
\begin{align}
\label{defV}
    \mathcal V \coloneqq \{ g_{\phi \phi} < 0 \} =  \left\{ r \leq 0, \, \Sigma \neq 0, \, \sin{\theta} \neq 0, \, \cos{(2\theta)} < - \dfrac{a^{4} + a^{2} r (2M + 3r) + 2 r^{4}}{a^{2} \Delta} \right\}.
\end{align}
The condition appearing in Eq.\ \eqref{defV} is a direct consequence of the expression
\begin{align}
    g_{\phi \phi} =  \dfrac{a^{4} + a^{2}\Delta \cos{(2 \theta)}  + a^{2} r (2M + 3r) + 2 r^{4}}{a^{2} \cos{(2 \theta)} + a^{2} + 2 r^{2}}\sin^{2}{\theta},
\end{align}
equivalent to Eq.\ \eqref{gff}. From condition (\ref{defV}) one obtains
\begin{align}
    1 \geq - \cos{(2 \theta)} > \dfrac{a^{4} + a^{2} r (2 M + 3 r) + 2 r^{4}}{a^{2} \Delta}
\end{align}
and thus
\begin{align}
\label{defVprime}
    0 > r \left[ a^2 (2M + r) + r^{3} \right]
\end{align}
in $\mathcal{V}$. Consequently, we define
\begin{align}
    \mathcal V' \coloneqq \{ r \le 0, \,  r \left[ a^2 (2M + r) + r^{3} \right] < 0 \},
\end{align}
so that $\mathcal V \subset \mathcal V'$.

The discriminant of the polynomial $\mathcal{P}(r) = r^{3} + a^{2}(r+2M)$ reads $-4 a^{6} - 108 a^{4} M^{2}$ and is clearly negative, as long as $a \neq 0$. Consequently, $\mathcal{P}(r)$ has only one real root, which we denote by $\hat{r}$. Since $\mathcal{P}(0) = 2 M a^2$ remains positive for $a \neq 0$ and $\lim_{r \to -\infty} \mathcal P(r) = -\infty$, $\hat r$ must be negative \footnote{In fact, $\hat{r}$ cannot be larger than $-|a|$, since $\mathcal{P}(-|a|) = 2|a|^{2} (M - |a|) \ge 0$ for $|a|\leq M$.}. It is given by
\begin{align}
    \hat{r}:=\sqrt[3]{\dfrac{\sqrt{3}\sqrt{a^{6}+27M^{2}a^{4}}-9Ma^{2}}{9}}-\dfrac{a^{2}}{\sqrt[3]{3}\sqrt[3]{\sqrt{3}\sqrt{a^{6}+27M^{2}a^{4}}-9Ma^{2}}}.
\end{align}
Note that $\hat{r}\to 0$, as $a\to 0$. The set $\mathcal V'$ can be simply characterized as $\mathcal V' = \{ \hat r < r < 0 \}$. 

\begin{figure}
    \centering
    \makebox[\textwidth][c]{%
    \hspace{-37.5pt}%
        \makebox[0pt][l]{%
            \hspace{2\dimexpr 0.33\textwidth - 24pt\relax}%
            \begin{minipage}{0.33\textwidth}
                \centering
                \includegraphics[width=\linewidth]{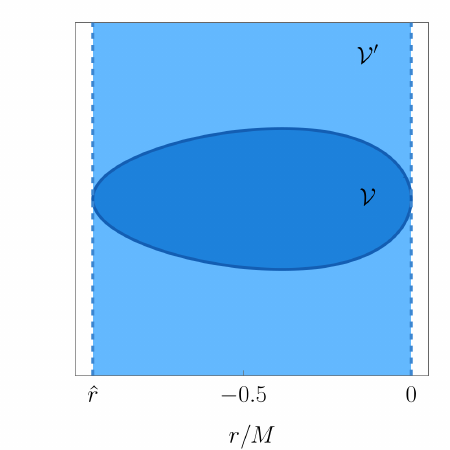}
                \vspace{0pt}
                \hspace{18.75pt}%
                (c)
            \end{minipage}%
        }%
        \makebox[0pt][l]{%
            \hspace{1\dimexpr 0.33\textwidth - 24pt\relax}%
            \begin{minipage}{0.33\textwidth}
                \centering
                \includegraphics[width=\linewidth]{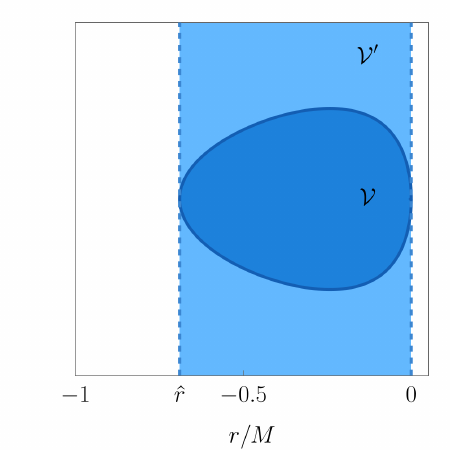}
                \vspace{0pt}
                \hspace{18.75pt}%
                (b)
            \end{minipage}%
        }%
        \begin{minipage}{0.33\textwidth}
            \centering
            \includegraphics[width=\linewidth]{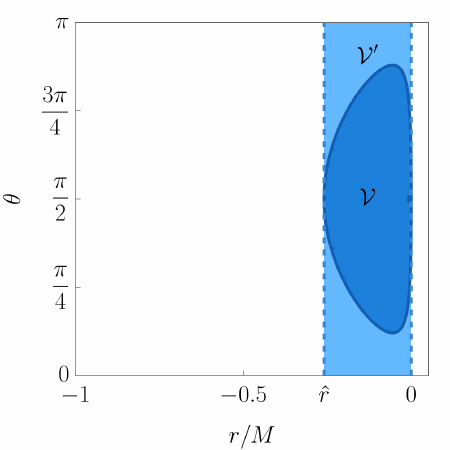}
            \vspace{0pt}
            \hspace{18.75pt}%
            (a)
        \end{minipage}%
        \hspace{2\dimexpr 0.33\textwidth - 30pt\relax}%
    }%
    \caption{\label{fig:forbid_regio} Regions $\mathcal{V}$ and $\mathcal{V}$' in the $(r,\theta)$ space. We set $M = 1$. Panel (a): $a = 0.1$, $\hat{r} = -0.259170$. Panel (b): $a = 0.5$, $\hat{r} = -0.689398$. Panel (c): $a = 0.9$, $\hat{r} = -0.948042$.}
\end{figure}

Both sets $\mathcal V$ and $\mathcal V'$ are plotted in Fig.\ \ref{fig:forbid_regio}. Note that the boundaries of the CTM set $\mathcal V$ are tangent to the boundaries of the cover set $\mathcal V'$ in the equatorial plane $\theta = \pi/2$. In this sense, the cover set $\mathcal V'$ provides an optimal characterization of $\mathcal V$ in the radial direction. On the other hand, the extent of $\mathcal V$ is bounded in the $\theta$ direction. For any $a$ such that $0 < |a| < M$, there exist $\theta_0 \in (0,\pi/2)$ such that $\mathcal V$ corresponds to points with $\theta \in [\theta_0, \pi - \theta_0]$. Anticipating the description of COS diagrams for the Kerr spacetime, provided in Sec.\ \ref{sec:projdiag}, we note that the set $\mathcal V'$ will always have to be excluded from the COS diagram, even though for the hypersurfaces of constant $\theta$ and $\tilde \phi$ (say) one can construct standard conformal diagrams that do not involve the CTM set $\mathcal V$, provided that $\theta$ is chosen sufficiently close to the axis $\theta \in \{ 0, \pi\}$.

\section{Projection diagrams}
\label{sec:projdiag}

\subsection{Effective metric}

The COS diagrams, introduced in \cite{COS2012}, are defined as follows. Let $(\mathcal{M}, g)$ be a smooth spacetime, and let $\mathcal{W}\subset\mathbb{R}^{1,1}$ be a subset of a two-dimensional Minkowski space $\mathbb{R}^{1,1}$. The pair $(\pi, \mathcal{U})$, where $\pi$ denotes a map $\pi\colon\mathcal{M}\to\mathcal{W}$ and $\mathcal{U}\subset\mathcal{M}$ is an open non-empty set in $\mathcal{M}$, is called a projection diagram if the following conditions are satisfied:
\begin{enumerate}
    \item The map $\pi$ is continuous and differentiable on an open dense set, and it is a smooth submersion on $\mathcal{U}$
    \item Every smooth timelike curve $\sigma\subset\pi(\mathcal{U})$ is the projection of a smooth timelike curve $\gamma$ in $(\mathcal{U},g)$: $\sigma=\pi\circ\gamma$
    \item The image $\pi\circ\gamma$ of every smooth timelike curve $\gamma\subset\mathcal{U}$ is a timelike curve in $\mathbb{R}^{1,1}$
\end{enumerate}

Note that the set $\mathcal{U}$ is effectively assumed to be of the same dimension as $\mathcal{M}$, which constitutes a significant advantage of COS diagrams over the traditional Carter-Penrose approach, where the corresponding hypersurface is restricted to two dimensions. Many subtle aspects of this definition are discussed in \cite{COS2012}. Here we only mention the obvious fact that the restriction to a subset $\mathcal U \subset \mathcal M$ allows one to remove a problematic region from the spacetime manifold, such as the CTM set in the extended Kerr spacetime.

We will now review the construction of COS diagrams for the Kerr spacetime. In Boyer-Lindquist coordinates, the map $\pi$ is defined as $\pi\colon (t,r,\theta,\phi)\to (t,r)$, which can later be composed with additional two-dimensional conformal transformations.

In the region in which the vector $\partial_{\phi}$ is spacelike, metric \eqref{KerrBL} takes the form
\begin{align}        
g&=g_{tt}^{\mathrm{eff}}dt^{2}+g_{\phi\phi}\left(d\phi+\dfrac{g_{t\phi}}{g_{\phi\phi}}dt\right)^{2}+g_{rr}dr^{2}+g_{\theta\theta}d\theta^{2},
\end{align}
where the effective component $g_{tt}^{\mathrm{eff}}$ is defined as
\begin{align}
g_{tt}^{\mathrm{eff}}(r,\theta)\coloneqq g_{tt} - \frac{g_{t\phi}^2}{g_{\phi\phi}} =-\dfrac{2\Sigma\Delta}{a^{4}+a^{2}\Delta\cos{(2\theta)}+a^{2}r(2M+3r)+2r^{4}}.
\end{align}
In the region $r\in \left(0,r_{-}\right)\cup\left(r_{+},\infty\right)$, this quantity can be bounded by
\begin{align}
\dfrac{\Sigma\Delta}{\left(r^{2}+a^{2}\right)^{2}}\leq\left|g_{tt}^{\mathrm{eff}}(r,\theta)\right|\leq\dfrac{\Sigma\Delta}{r\left[r^{3}+\left(r+2M\right)a^{2}\right]},
\label{g00eff_bound}
\end{align}
which follows directly from estimating the trigonometric term in the denominator as $-1\le\cos{(2 \theta)}\le 1$. Both the lower and upper bounds depend on $\theta$ via $\Sigma$. Consequently, they attain a minimum at $\theta = \pi/2$ and a maximum at $\theta = 0$ and $\theta = \pi$. A careful inspection shows that the term $\left|g_{tt}^{\mathrm{eff}}(r,\theta)\right|$ also attains a minimum with respect to $\theta$ at $\theta = \pi/2$.

Consider a vector $u$ assumed to be causal with respect to the metric $g$, i.e., satisfying the condition
\begin{align}
\label{causal1}
    0 \geq g(u,u) = g_{tt}^\mathrm{eff} \left( u^{t} \right)^{2} + g_{\phi\phi} \left( u^{\phi} + \dfrac{g_{t\phi}}{g_{\phi\phi}} u^{t} \right)^{2} + g_{rr} \left( u^{r} \right)^{2} + g_{\theta\theta} \left( u^{\theta} \right)^{2}.
\end{align}
Let $\Omega^2(r,\theta)$ be any positive function of $r$ and $\theta$. Equation (\ref{causal1}) implies that for $r \in \left(0,r_{-}\right)\cup\left(r_{+},\infty\right)$ (in particular, outside the CTM set), one has
\begin{align}
\label{estimate1}
    0 \geq \Omega^{2}(r,\theta) g(u,u) \geq \Omega^{2}(r,\theta) \left[ g_{tt}^\mathrm{eff} \left( u^{t} \right)^{2} + g_{rr} \left( u^{r} \right)^{2} \right] \geq -\sup_{\theta \in [0,\pi]}\left(\Omega^{2}\left|g_{tt}^{\mathrm{eff}}\right|\right)\left(u^{t}\right)^{2}+\inf_{\theta \in [0,\pi]}\left(\Omega^{2}g_{rr}\right)\left(u^{r}\right)^{2}.
\end{align}
Following \cite{COS2012}, we now take
\begin{align}
\label{defOmega}
\Omega^{2}=\dfrac{r^{2}+a^{2}}{\Sigma}.
\end{align}
In the region $r \in \left(0,r_{-}\right)\cup\left(r_{+},\infty\right)$ this choice implies that
\begin{align}
\label{cond2}
    0 \geq - \dfrac{\left( r^{2} + a^{2} \right)\Delta}{r \left[ r^{3} + (r+2M)a^{2} \right]} \left(u^{t}\right)^{2} + \dfrac{r^{2} + a^{2}}{\Delta} \left( u^{r} \right)^{2},
\end{align}
where we have used the bound (\ref{g00eff_bound}). Condition (\ref{cond2}) can be reformulated as a causality condition with respect to the two-dimensional metric
\begin{align}
    \nonumber 
        g_{p}&=-\dfrac{\left(r^{2}+a^{2}\right)\Delta}{r\left[r^{3}+\left(r+2M\right)a^{2}\right]}dt^{2}+\dfrac{r^{2}+a^{2}}{\Delta}dr^{2} \\
        &=\dfrac{r^{2}+a^{2}}{\sqrt{r\left[r^{3}+\left(r+2M\right)a^{2}\right]}}\left[-\dfrac{\Delta}{\sqrt{r\left[r^{3}+\left(r+2M\right)a^{2}\right]}}dt^{2}+\dfrac{\sqrt{r\left[r^{3}+\left(r+2M\right)a^{2}\right]}}{\Delta}dr^{2}\right].
        \label{proj_metric}
\end{align}
More precisely, Eq.\ \eqref{cond2} can be written as $0 \geq g_p(\pi_\ast(u),\pi_\ast(u))$, where $\pi_\ast(u) = u^t \partial_t + u^r \partial_r$ denotes the pushforward of $u$ by $\pi$.

In the range $r \in (r_-,r_+)$, i.e., for $\Delta < 0$, we have, instead of Eq.\ \eqref{g00eff_bound},
\begin{align}
- \frac{\Sigma\Delta}{r\left[r^{3}+\left(r+2M\right)a^{2}\right]}
\leq\left|g_{tt}^{\mathrm{eff}}(r,\theta)\right| \leq - \frac{\Sigma\Delta}{\left(r^{2}+a^{2}\right)^{2}}.
\label{g00eff_bound2}
\end{align}
Similarly to Eq.\ \eqref{estimate1}, we now estimate
\begin{align}
\label{estimate2}
    0 \ge  \Omega^2(r,\theta) \left[ g_{tt}^\mathrm{eff} \left( u^t \right)^2 + g_{rr} \left( u^r \right)^2 \right] \ge \inf_{\theta \in [0,\pi]}\left(\Omega^{2}\left|g_{tt}^{\mathrm{eff}}\right|\right)\left(u^t \right)^2 - \sup_{\theta \in [0,\pi]}\left(\Omega^2 \left| g_{rr} \right| \right) \left( u^r \right)^2. 
\end{align}
Choosing again $\Omega^2$ given by Eq.\ \eqref{defOmega} and using inequality (\ref{estimate2}), we see that Eq.\ \eqref{cond2} also holds for $r \in (r_-,r_+)$. Consequently, the relation $0 \ge g_p(\pi_{\ast}(u),\pi_{\ast}(u))$ is satisfied for all $r > 0$.

For $r<0$, Eq.\ \eqref{causal1}  holds for all points outside the CTM set. Moreover, outside the set $\mathcal{V}'$, one has $r \left[ a^{2} (2M + r) + r^{3} \right] > 0$, and Eq.\ \eqref{g00eff_bound} is also satisfied. This implies that for $r < 0$ and outside $\mathcal{V}'$ Eq.\ \eqref{cond2} is valid. In summary, we conclude that the condition $0 \geq g_p(\pi_{\ast}(u),\pi_{\ast}(u))$ holds in the entire range of $r$, except for the set $\mathcal{V}'$.

We emphasize that the advantage of the construction described up to this point is that neither the metric $g_p$ nor the characterization of the set $\mathcal{V}'$ depend on the polar angle $\theta$. One should also stress that due to the choice of the conformal factor $\Omega^2$ in Eq.\ \eqref{defOmega}, the polynomial $r[r^3 + (r + 2 M)a^2]$ appearing in the expression for the metric $g_p$ coincides with the same polynomial, defining the set $\mathcal V'$.

Subsequent steps of the construction are quite standard. The projection $\pi$ can be composed with standard Kruskal-Szekeres-type transformations, used in the usual construction of Carter-Penrose conformal diagrams. These transformations can be applied directly to the metric $g_p$, which we express as
\begin{align}
    g_p = \dfrac{r^{2}+a^{2}}{\sqrt{r\left(r^{3}+\left(r+2M\right)a^{2}\right)}} \left[ -F(r) dt^{2} + F(r)^{-1} dr^{2} \right],
\end{align}
where
\begin{align}
    F(r)=\dfrac{\Delta}{\sqrt{r\left(r^{3}+\left(r+2M\right)a^{2}\right)}}.
\end{align}
In contrast to many text-book cases, the integral $\int F(r)^{-1} dr$, appearing in this construction, cannot be easily evaluated analytically, and one has to resort to numerical methods. This requires some subtle treatment related to the gluing of individual conformal blocks, which we describe in the following subsection.

\subsection{Null coordinates and the gluing procedure}

We will now describe the construction of individual conformal blocks and the gluing procedure \`{a} la Kruskal and Szekeres \cite{Kruskal1960,Szekeres}. It is based on a general scheme discussed in \cite{W1970} and also used in \cite{COS2012}. Since in this work, we aim at an exact numerical description, instead of schematic drawings, we will provide a little more details than usual.

In the first step, one introduces the advanced and retarded null coordinates
\begin{subequations}
\begin{align}
    v&=t+f(r), \\
    w&=t-f(r),
\end{align}
\end{subequations}
where $f'(r) = F(r)^{-1}$. In each of the four regions $r\in\left(r_{+},\infty\right)$, $r\in\left(r_{-},r_{+}\right)$, $r\in\left(0,r_{-}\right)$, and $r \in (-\infty,\hat r)$, the function $f(r)$ can be expressed as
\begin{align}
\label{deff}
    f(r) = \int\limits_{r_i}^{r} \frac{d \bar r}{F(\bar r)} + f_{0,i}, \quad i = 1, \dots, 4 ,
\end{align}
where $r_{1} \in (r_+,\infty)$, $r_{2} \in (r_{-},r_{+})$, $r_{3} \in (0,r_{-})$, $r_4 \in (-\infty,\hat{r})$, and $f_{0,i}$ denote arbitrary constants. In principle, introducing constants $f_{0,i}$ is redundant, as the exact numerical value of $f(r)$ can be controlled by the choice of the reference radius $r_i$. Keeping the constants $f_{0,i}$ separately turns out to be convenient in the numerical implementation. In what follows, we will refer to regions $r\in\left(r_{+},\infty\right)$, $r\in\left(r_{-},r_{+}\right)$, $r\in\left(0,r_{-}\right)$, and $r \in (-\infty,\hat r)$, as Region 1, \dots, Region 4, respectively. The behavior of $f(r)$ in Regions 1 and 2 is relatively simple. Since $F(r)$ has simple zeros at $r = r_\pm$, $f(r)$ diverges at the horizons. In Region 1, $f(r)$ goes from $-\infty$ to $+\infty$, as $r$ changes from $r_+$ to $+\infty$. In Region 2, $f(r)$ decreases from $+\infty$ to $-\infty$, as $r$ changes from $r_{-}$ to $r_{+}$. Regions 3 and 4 cover the range $r \in (-\infty,r_{-})$, except for the set $\mathcal{V}'$, in which $F(r)$ is imaginary. In Region 3, $f(r)$ increases from
\begin{align}
    \int\limits_{r_{3}}^{0} \dfrac{d\bar{r}}{F(\bar{r})} + f_{0,3}
\end{align}
to $+\infty$, as $r$ changes from $0$ to $r_{-}$. In Region 4, $f(r)$ changes from $-\infty$ to 
\begin{align}
    \int\limits_{r_4}^{\hat{r}} \dfrac{d\bar{r}}{F(\bar r)} + f_{0,4},
\end{align}
as $r$ increases from $-\infty$ to $\hat{r}$. In all regions, $t\in\mathbb{R}$.

In terms of the coordinates $v$ and $w$, the line element $-F(r) dt^{2} + F(r)^{-1}dr^{2}$ can be written as
\begin{align}
\label{lineelement}
    -F(r) dt^{2} + F(r)^{-1}dr^{2} = -F(r) dv dw,
\end{align}
and consequently,
\begin{align}
\label{gpnew}
    g_p = -\dfrac{(r^{2}+a^{2})\Delta}{r\left(r^{3}+\left(r+2M\right)a^{2}\right)} dv dw,
\end{align}
which is a well-known textbook fact. The presence of the conformal factor $F(r)$ in Eq.\ \eqref{lineelement}, or $\Delta$ in the conformal factor in Eq.\ \eqref{gpnew}, creates difficulty, as these terms vanish at the horizons $r = r_{\pm}$. This problem can be remedied by a simple exponential transformation
\begin{subequations}
\label{tildetransf}
\begin{align}
    \tilde{v}&=+e^{+kv}, \\
    \tilde{w}&=-e^{-kw},
\end{align}
\end{subequations}
so that
\begin{align}
    g_{p}=-\dfrac{\left(r^{2}+a^{2}\right)\Delta}{k^{2}r\left(r^{3}+\left(r+2M\right)a^{2}\right)}e^{-2kf(r)}d\tilde{v}d\tilde{w},
\label{metric_KS_cords}
\end{align}
where $k$ is a constant. The choice of $k$ is a part of gluing between regions 1, 2, and 3, and it can be described as follows.

We start by discussing the freedom of choosing reference radii $r_{i}$ and the constants $f_{0,i}$ in Eq.\ \eqref{deff}. For clarity, consider a gluing between Regions 1 and 2. Fix the radii $r_1 \in (r_{+},+\infty)$ and $r_2 \in (r_{-},r_{+})$. Smooth gluing between Regions 1 and 2 can be achieved by an appropriate choice of constants $f_{0,1}$ and $f_{0,2}$. Since $F(r)$ has a simple zero at $r = r_{+}$, one can write
\begin{align}
    F(r)=\left(r-r_{+}\right)H(r),
\end{align}
where
\begin{align}
    H(r) = \dfrac{r - r_{-}}{\sqrt{r\left(r^{3}+\left(r+2M\right)a^{2}\right)}}
\end{align}
has no zeros in a neighborhood of $r = r_{+}$, and in fact in $(r_-,+\infty)$. We require that the numerical value of the function
\begin{align}
    \hat{f}(r) \coloneqq f(r)-\dfrac{1}{H(r_{+})}\log{\left|\dfrac{r-r_{+}}{r_{+}}\right|}
\end{align}
remain continuous across the horizon $r = r_{+}$. One-sided limits of $\hat{f}(r)$ at $r = r_{+}$ read
\begin{subequations}
    \begin{align}
    \lim_{r \searrow r_{+}} \hat{f}(r) & = \int\limits_{r_{1}}^{r_{+}} \left[ \dfrac{d\bar{r}}{\bar{r} - r_{+}} \left( \dfrac{1}{H(\bar{r})} - \dfrac{1}{H(r_{+})} \right) \right] - \dfrac{1}{H(r_{+})} \log \left| \dfrac{r_{1} - r_{+}}{r_{+}} \right| + f_{0,1}, 
    \\
    \lim_{r \nearrow r_{+}} \hat{f}(r) & = \int\limits_{r_{2}}^{r_{+}} \left[ \frac{d\bar{r}}{\bar{r} - r_{+}} \left( \dfrac{1}{H(\bar{r})} - \dfrac{1}{H(r_{+})} \right) \right] - \dfrac{1}{H(r_{+})} \log \left| \dfrac{r_{2} - r_{+}}{r_{+}} \right| + f_{0,2}. 
\end{align}
\end{subequations}
Consequently, the condition $\lim\limits_{r \searrow r_{+}} \hat{f}(r) = \lim\limits_{r \nearrow r_{+}} \hat{f}(r)$ implies that
 \begin{align}
    \nonumber 
     f_{0,1} - f_{0,2}  = & \int\limits_{r_{2}}^{r_{+}} \left[ \dfrac{d\bar{r}}{\bar{r} - r_{+}} \left( \dfrac{1}{H(\bar{r})} - \dfrac{1}{H(r_{+})} \right) \right] - \dfrac{1}{H(r_{+})} \log \left| \dfrac{r_{2} - r_{+}}{r_{+}} \right| \\
     & - \int\limits_{r_{1}}^{r_{+}} \left[ \dfrac{d\bar{r}}{\bar{r} - r_{+}} \left( \dfrac{1}{H(\bar{r})} - \dfrac{1}{H(r_{+})} \right) \right] + \dfrac{1}{H(r_{+})} \log \left| \frac{r_{1} - r_{+}}{r_+} \right|.
     \label{constdiff}
 \end{align}
Note that all terms appearing in Eq.\ \eqref{constdiff} are regular. In particular, the integrand
\begin{align}
    \dfrac{1}{r - r_{+}} \left( \dfrac{1}{H(r)} - \dfrac{1}{H(r_{+})} \right)= \dfrac{H(r_{+}) - H(r)}{(r - r_{+}) H(r_{+}) H(r)}
\end{align}
has a limit
\begin{align}
    \lim_{r \to r_{+}} \dfrac{1}{r - r_{+}} \left( \dfrac{1}{H(r)} - \dfrac{1}{H(r_{+})} \right) = \dfrac{r_{+}^{3}(r_{+} - 2 r_{-}) - a^{2} \left[r_{+} r_{-} + M (r_{+} + r_{-})\right]}{(r_{+} - r_{-})^{2} \sqrt{r_{+} \left[r_{+}^{3} + (r_{+} + 2M) a^{2}\right]}},
\end{align}
which remains finite, as long as $|a| < M$.
An analogous condition has to be imposed while gluing Regions 2 and 3 at $r = r_{-}$. This yields, in total, 2 conditions that involve 3 constants $f_{0,1}$, $f_{0,2}$, and $f_{0,3}$. The constant $f_{0,4}$ can be specified freely, as there is no gluing between Regions 3 and 4.

Returning to the choice of the constant $k$ in Eq.\ \eqref{metric_KS_cords}, we note that a regular gluing between Regions 1 and 2 can be achieved 
by selecting
\begin{align}
    k = k_{+} \coloneqq \dfrac{1}{2}\dfrac{dF}{dr}\Bigg|_{r=r_+}.
\end{align}
Analogously, regular gluing between Regions 2 and 3 is obtained with
\begin{align}
    k = k_{-} \coloneqq \dfrac{1}{2}\dfrac{dF}{dr}\Bigg|_{r=r_-}.
\end{align}
In the case of the gluing between Regions 1 and 2, one has $k_{+} = \frac{1}{2}H(r_{+})$, and consequently,
\begin{align}
    e^{-2 k_{+} f(r)} = \left| \dfrac{r_{+}}{r - r_{+}} \right| e^{-H(r_{+}) \hat{f}(r)}.
\end{align}
As a result, metric \eqref{metric_KS_cords} can be written as
\begin{align}
g_{p}=\mp\dfrac{r_{+}\left(r-r_{-}\right)\left(r^{2}+a^{2}\right)}{k_{+}^{2}r\left(r^{3}+\left(r+2M\right)a^{2}\right)}e^{-2k_{+}\hat{f}(r)}d\tilde{v}d\tilde{w},
\end{align}
where the minus sign corresponds to Region 1, whereas the plus sign corresponds to Region 2. Similarly, imposing the regularity condition at the Cauchy horizon $r=r_{-}$ yields the metric
\begin{align}
    g_{p}=\mp\dfrac{r_{-}\left(r-r_{+}\right)\left(r^{2}+a^{2}\right)}{k_{-}^{2}r\left(r^{3}+\left(r+2M\right)a^{2}\right)}e^{-2k_{-}\hat{f}(r)}d\tilde{v}d\tilde{w},
\end{align}
with the minus and plus signs corresponding to Regions 2 and 3, respectively. In this formulation, $r=r\left(\tilde{v},\tilde{w}\right)$ satisfies the relation $2kf(r)=\log{\left|-\tilde{v}\tilde{w}\right|}$, which can be numerically inverted.

\subsection{Compactification, conformal blocks}

\renewcommand{\arraystretch}{1.5}
\begin{table}[t]
\begin{ruledtabular}
\caption{\label{tab:TXcoor} Compactified coordinates $(T,X)$ in Blocks I--III and I$'$--III$'$.}
    \begin{tabular}{ccc}
          $k$ & $k_{+}$ & $k_{-}$ \\
         \colrule
          \textbf{Block I} & $\rule{0pt}{2.66em}\begin{aligned}T&=\hphantom{+}\dfrac{1}{2}\left(V+W\right)\\X&=\hphantom{+}\dfrac{1}{2}\left(V-W\right)\end{aligned}\rule[-2.33em]{0pt}{0pt}$ & \\
          \colrule
          \textbf{Block I$'$} & $\rule{0pt}{2.66em}\begin{aligned}T&=\hphantom{+}\dfrac{1}{2}\left(V+W\right)\\X&=\hphantom{+}\dfrac{1}{2}\left(V-W\right)\end{aligned}\rule[-2.33em]{0pt}{0pt}$ & \\
         \colrule
          \textbf{Block II} & $\rule{0pt}{2.66em}\begin{aligned}T&=\hphantom{+}\dfrac{1}{2}\left(V-W\right)\\X&=\hphantom{+}\dfrac{1}{2}\left(V+W\right)\end{aligned}\rule[-2.33em]{0pt}{0pt}$ & $\rule{0pt}{2.66em}\begin{aligned}T&=-\dfrac{1}{2}\left(V-W\right)+\dfrac{\pi}{2}\\X&=-\dfrac{1}{2}\left(V+W\right)\end{aligned}\rule[-2.33em]{0pt}{0pt}$ \\
         \colrule
          \textbf{Block II$'$} & $\rule{0pt}{2.66em}\begin{aligned}T&=-\dfrac{1}{2}\left(V-W\right)\\X&=-\dfrac{1}{2}\left(V+W\right)\end{aligned}\rule[-2.33em]{0pt}{0pt}$ & $\rule{0pt}{2.66em}\begin{aligned}T&=\hphantom{+}\dfrac{1}{2}\left(V-W\right)+\dfrac{\pi}{2}\\X&=\hphantom{+}\dfrac{1}{2}\left(V+W\right)\end{aligned}\rule[-2.33em]{0pt}{0pt}$ \\
         \colrule
          \textbf{Block III} & & $\rule{0pt}{2.66em}\begin{aligned}T&=-\dfrac{1}{2}\left(V+W\right)+\dfrac{\pi}{2}\\X&=-\dfrac{1}{2}\left(V-W\right)\end{aligned}\rule[-2.33em]{0pt}{0pt}$\\
         \colrule
          \textbf{Block III$'$} & & $\rule{0pt}{2.66em}\begin{aligned}T&=-\dfrac{1}{2}\left(V+W\right)+\dfrac{\pi}{2}\\X&=-\dfrac{1}{2}\left(V-W\right)\end{aligned}\rule[-2.33em]{0pt}{0pt}$
    \end{tabular}
\end{ruledtabular}
\end{table}

\begin{figure}[t]
    \centering
    \begin{minipage}[c]{0.35\textwidth}
    \centering
    \includegraphics[width=\textwidth]{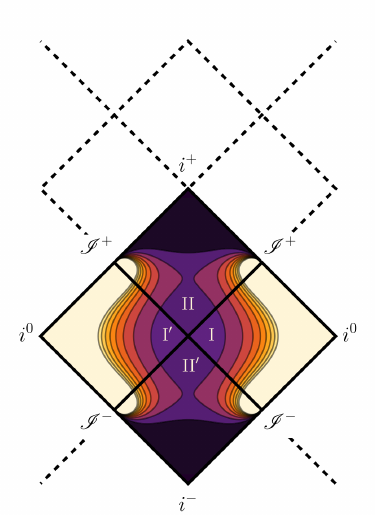}
    \begin{tikzpicture}[remember picture, overlay]
            \node (ref) at (0,0) {}; 
        \end{tikzpicture}
  \end{minipage}
  \hspace{0.1\textwidth}
  \begin{minipage}[c]{0.35\textwidth}
    \centering
    \includegraphics[width=\textwidth]{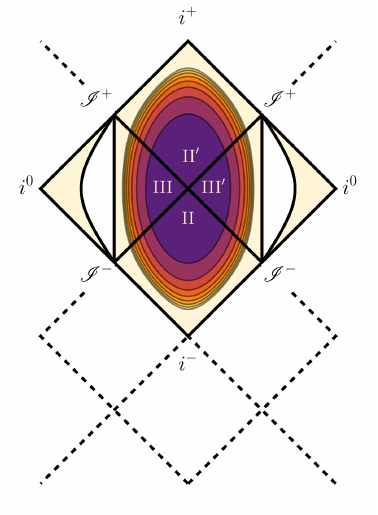}
    \begin{tikzpicture}[remember picture, overlay]
            \node (ref2) at (0,0) {}; 
        \end{tikzpicture}
  \end{minipage}

    \begin{tikzpicture}[remember picture, overlay]
        \node (colorbar) at ([xshift=3.75cm, yshift=4cm]ref) 
            {\includegraphics[width=0.055\textwidth]{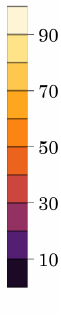}};
        \node[above=0.0cm of colorbar,xshift=-0.125cm, font=\large] {$\Phi^{2}$};
    \end{tikzpicture}

    \begin{tikzpicture}[remember picture, overlay]
        \node (colorbar2) at ([xshift=3.75cm, yshift=4cm]ref2) 
            {\includegraphics[width=0.045\textwidth]{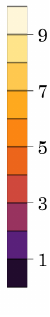}};
        \node[above=0.0cm of colorbar2,xshift=-0.05cm, font=\large] {$\Psi^{2}$};
    \end{tikzpicture}
  
    \caption{\label{fig:conffactors} Conformal factors $\Phi^{2}(T,X)$ and $\Psi^{2}(T,X)$ for $M=1$ and $a=0.9$. We denote by $i^{-}$, $i^{0}$, and $i^{+}$ the past timelike infinity, spatial infinity, and future timelike infinity, respectively. The symbols $\mathscr{I}^{-}$ and $\mathscr{I}^{+}$ denote the past and future null infinity.}
\end{figure}

Compactification of null coordinates $\tilde v$ and $\tilde w$ can be obtained in the standard way, by defining
\begin{subequations}
\begin{align}
    V& = \arctan{\tilde{v}}, \\
    W& = \arctan{\tilde{w}},
\end{align}
\end{subequations}
which gives rise to conformal Blocks I--III and their copies I$'$--III$'$, covering the maximally extended Kerr spacetime and constituting the resulting COS diagram, as illustrated in Fig.\ \ref{fig:conffactors}. Our numbering of blocks follows that of \cite{Neill1995}. Blocks I and I$'$ correspond to Region 1, while Blocks II and II$'$ correspond to Region 2. Each of Blocks III and III$'$ consists of Regions 3 and 4, and contains the set $\mathcal V'$. Blocks I, II, I$'$ and II$'$ can be smoothly joined along the boundaries corresponding to $r = r_+$. Similarly, Blocks II, III, II$'$, and III$'$ can be joined along the boundaries corresponding to $r = r_-$. This can be achieved by introducing global time and spatial coordinates $T$ and $X$, which we define according to Table \ref{tab:TXcoor}. As described in the preceding subsection, smooth joining along $r = r_\pm$ requires setting $k = k_\pm$, respectively. Consequently, the transformation $(t,f(r))\to(T, X)$ reads, for $k=k_{+}$,
\begin{subequations}
    \begin{align}
        T&=\dfrac{1}{2}\left[\arctan{\left(e^{k(t+f)}\right)}+\arctan{\left(-e^{-k(t-f)}\right)}\right], \\
        X&=\dfrac{1}{2}\left[\arctan{\left(e^{k(t+f)}\right)}-\arctan{\left(-e^{-k(t-f)}\right)}\right]
    \end{align}
\end{subequations}
in Block I and Block I$'$,
\begin{subequations}
    \begin{align}
        T&=\dfrac{1}{2}\left[\arctan{\left(e^{k(t+f)}\right)}-\arctan{\left(-e^{-k(t-f)}\right)}\right], \\
        X&=\dfrac{1}{2}\left[\arctan{\left(e^{k(t+f)}\right)}+\arctan{\left(-e^{-k(t-f)}\right)}\right]
    \end{align}
\end{subequations}
in Block II and
\begin{subequations}
    \begin{align}
        T&=-\dfrac{1}{2}\left[\arctan{\left(e^{k(t+f)}\right)}-\arctan{\left(-e^{-k(t-f)}\right)}\right], \\
        X&=-\dfrac{1}{2}\left[\arctan{\left(e^{k(t+f)}\right)}+\arctan{\left(-e^{-k(t-f)}\right)}\right]
    \end{align}
\end{subequations}
in Block II$'$. Similarly, for $k=k_{-}$ one obtains the formulas
\begin{subequations}
    \begin{align}
        T&=-\dfrac{1}{2}\left[\arctan{\left(e^{k(t+f)}\right)}-\arctan{\left(-e^{-k(t-f)}\right)}\right]+\dfrac{\pi}{2}, \\
        X&=-\dfrac{1}{2}\left[\arctan{\left(e^{k(t+f)}\right)}+\arctan{\left(-e^{-k(t-f)}\right)}\right]
    \end{align}
\end{subequations}
in Block II,
\begin{subequations}
    \begin{align}
        T&=\dfrac{1}{2}\left[\arctan{\left(e^{k(t+f)}\right)}-\arctan{\left(-e^{-k(t-f)}\right)}\right]+\dfrac{\pi}{2}, \\
        X&=\dfrac{1}{2}\left[\arctan{\left(e^{k(t+f)}\right)}+\arctan{\left(-e^{-k(t-f)}\right)}\right]
    \end{align}
\end{subequations}
in Block II$'$, and
\begin{subequations}
    \begin{align}
        T&=-\dfrac{1}{2}\left[\arctan{\left(e^{k(t+f)}\right)}+\arctan{\left(-e^{-k(t-f)}\right)}\right]+\dfrac{\pi}{2}, \\
        X&=-\dfrac{1}{2}\left[\arctan{\left(e^{k(t+f)}\right)}-\arctan{\left(-e^{-k(t-f)}\right)}\right]
    \end{align}
\end{subequations}
in Block III and Block III$'$. One should also emphasize that in the diagram corresponding to the maximally extended Kerr spacetime, all blocks are repeated cyclically in $T$. The definitions collected in Table \ref{tab:TXcoor} refer to a single copy of each of the blocks.

Setting $k = k_{+}$, one can express the metric \eqref{proj_metric} in terms of variables $T$ and $X$ as
\begin{align}
g_{p}=\underbrace{\dfrac{r_{+}\left(r-r_{-}\right)\left(r^{2}+a^{2}\right)}{k_{+}^{2}r\left(r^{3}+\left(r+2M\right)a^{2}\right)\cos^{2}{\left(T+X\right)}\cos^{2}{\left(T-X\right)}}e^{-2k_{+}\hat{f}(r)}}_{=\Phi^{2}\left(T,X\right)}\left(-dT^{2}+dX^{2}\right),
\end{align}
which is explicitly regular at the horizon $r = r_{+}$. This form extends smoothly over Blocks I, II, I$'$, and II$'$. The resulting conformal factor will be denoted as $\Phi^2$. The form of the metric \eqref{proj_metric} regular at $r=r_{-}$ can be obtained by setting $k = k_{-}$ as
\begin{align}
    g_{p}=\underbrace{-\dfrac{r_{-}\left(r-r_{+}\right)\left(r^{2}+a^{2}\right)}{k_{-}^{2}r\left(r^{3}+\left(r+2M\right)a^{2}\right)\cos^{2}{\left(T+X-\frac{\pi}{2}\right)}\cos^{2}{\left(T-X-\frac{\pi}{2}\right)}}e^{-2k_{-}\hat{f}(r)}}_{=\Psi^{2}(T,X)}\left(-dT^{2}+dX^{2}\right).
\end{align}
It allows us to smoothly connect Blocks II and III, as well as their copies II$'$ and III$'$. In this case, the overall conformal factor is denoted by $\Psi^2$. The sample graphs of the conformal factors $\Phi^2$ and $\Psi^2$ are shown in Fig.\ \ref{fig:conffactors}.

In constructing the conformal diagrams shown in Fig.\ \ref{fig:conffactors}, it is convenient to directly relate the coordinates $(t,f(r))$ and $(T,X)$. As before, the exact formulas differ from one block to another. For the graphs shown in Fig.\ \ref{fig:conffactors}, these relations can be found as follows. For the blocks joined along $r = r_{+}$ assuming $k = k_{+}$, one has
\begin{align}   \tanh{\left(kt\right)}=\dfrac{\sin{2T}}{\sin{2X}},\qquad\tanh{\left(kf\right)}=-\dfrac{\cos{2X}}{\cos{2T}}
\end{align}
in Region 1 (Blocks I and I$'$), and
\begin{align}
\tanh{\left(kt\right)}=\dfrac{\sin{2X}}{\sin{2T}},\qquad\tanh{\left(kf\right)}=-\dfrac{\cos{2T}}{\cos{2X}}
\end{align}
in Region 2 (Blocks II and II$'$). For the blocks joined along $r = r_-$ assuming $k = k_-$, the corresponding formulas read
\begin{align}
\tanh{\left(kt\right)}=-\dfrac{\sin{2X}}{\sin{2T}},\qquad\tanh{\left(kf\right)}=\dfrac{\cos{2T}}{\cos{2X}}
\end{align}
in Region 2 (Blocks II and II$'$), and
\begin{align}
\tanh{\left(kt\right)}=-\dfrac{\sin{2T}}{\sin{2X}},\qquad\tanh{\left(kf\right)}=\dfrac{\cos{2X}}{\cos{2T}}
\end{align}
in Regions 3 and 4 (Blocks III and III$'$).

We emphasize that the coordinates $(T,X)$ do not extend smoothly to more than four blocks, as the corresponding conformal factor vanishes or diverges on the respective boundaries and likewise tends to infinity along the lines representing $r=0$ and $r=\hat{r}$ \cite{COS2012,W1970}. Due to this fact, related directly to the Kruskal-Szekeres-type gluing, a depiction of a geodesic passing through different blocks may require more than one diagram. Notably, coordinate systems covering more causal blocks exist, e.g., for the Reissner-Nordstr\"{o}m spacetime \cite{Israel,Pajerski,Klosch} (see also \cite{Bisson} for a recent generalization). Whether such a coordinate system (usually referred to as Israel coordinates) can be generalized to the case of COS diagrams for the Kerr metric remains, to our knowledge, an open problem.

\section{Geodesic equations}
\label{sec:geoeq}

Our description of causal Kerr geodesics follows closely \cite{CMH2023,B2025}. In Boyer-Lindquist coordinates, causal geodesics are described by the following set of equations \cite{Neill1995,Chandrasekhar}:
\begin{subequations}
\label{geoeq1}
\begin{align}
\Sigma \dfrac{dt}{d\lambda} &= \dfrac{\left(r^{2}+a^{2}\right)\left[\left(r^{2}+a^{2}\right)E-al_{z}\right]}{\Delta}+a\left(l_{z}-aE\sin^{2}{\theta}\right), \\
\Sigma \dfrac{dr}{d\lambda} &= \epsilon_{r}\sqrt{R(r)}, \label{geoeq1rad} \\
\Sigma \dfrac{d\theta}{d\lambda} &= \epsilon_{\theta}\sqrt{\Theta(\theta)}, \label{geoeq1polar} \\
\Sigma \dfrac{d\phi}{d\lambda} &= \dfrac{a\left[\left(r^{2}+a^{2}\right)E-al_{z}\right]}{\Delta}+\dfrac{1}{\sin^{2}{\theta}}\left(l_{z}-aE\sin^{2}{\theta}\right),
\end{align}
\end{subequations}
where
\begin{subequations}
\begin{align}
R(r) & = \left[\left(r^{2}+a^{2}\right)E-al_{z}\right]^{2}-\Delta\left(\mathcal{K}+m^{2}r^{2}\right), \\
\Theta(\theta) &= \mathcal{K}-m^{2}a^{2}\cos^{2}{\theta}-\left(\dfrac{l_{z}}{\sin{\theta}}-aE\sin{\theta}\right)^{2}.
\end{align}
\end{subequations}
The quantities $m$, $E$, $l_z$, and $\mathcal{K}$ denote the four constants of motion: the rest mass of the particle, its energy, its angular momentum, and the Carter constant, respectively \cite{C1968}. The parameters $\epsilon_{r}$ and $\epsilon_{\theta}$ are equal to $+1$ or $-1$, corresponding to the direction of motion relative to the radial and polar coordinates. For timelike geodesics, the parameter $\lambda$ is related to the proper time $\tau$ by $\lambda = \tau/m $. The components of the four-momenta $p^\mu = d x^\mu/d \lambda$ are normalized as $g(p,p) = p_\mu p^\mu = - m^2$. The components of the four-velocity $u^{\mu}=dx^{\mu}/d\tau$ satisfy $g(u,u) = g_{\mu\nu}u^{\mu}u^{\nu}=-\delta$, where the parameter 
\begin{align}
\delta=\begin{cases}
1 & \text{for timelike geodesics,} \\
0 & \text{for null geodesics}
\end{cases}
\end{align}
has been introduced for convenience of switching between the descriptions of timelike and null geodesics. The energy $E$ and the angular momentum $l_z$ are defined as $E = -p_t$ and $l_z = p_\phi$.

Radial and polar equations \eqref{geoeq1rad} and \eqref{geoeq1polar} can be decoupled by a suitable reparameterization, introduced by Mino in \cite{M2003} and defined by
\begin{align}
    \dfrac{dx^{\mu}}{d\tilde{s}}=\Sigma\dfrac{dx^{\mu}}{d\lambda}.
\end{align}
The parameter $\tilde s$ is usually referred to as the Mino time. In terms of $\tilde s$, Eqs.\ \eqref{geoeq1} can be written as
\begin{subequations}
\label{geoeq2}
\begin{align}
    \dfrac{dt}{d\tilde{s}}&=\dfrac{\left(r^{2}+a^{2}\right)\left[\left(r^{2}+a^{2}\right)E-al_{z}\right]}{\Delta}+a\left(l_{z}-aE\sin^{2}{\theta}\right), \\
    \dfrac{dr}{d\tilde{s}}&=\epsilon_{r}\sqrt{R(r)}, \\
    \dfrac{d\theta}{d\tilde{s}}&=\epsilon_{\theta}\sqrt{\Theta(\theta)}, \\
    \dfrac{d\phi}{d\tilde{s}}&=\dfrac{a\left[\left(r^{2}+a^{2}\right)E-al_{z}\right]}{\Delta}+\dfrac{1}{\sin^{2}{\theta}}\left(l_{z}-aE\sin^{2}{\theta}\right).    
\end{align}
\end{subequations}
Equations \eqref{geoeq2} can easily be transformed to the horizon-penetrating coordinates defined by  Eqs.\ \eqref{horizonpenetrating}. 
Upon this transformation, the time and azimuthal equations become
\begin{subequations}
\begin{align}
    \dfrac{d\tilde{t}}{d\tilde{s}}&=\dfrac{\left(r^{2}+a^{2}\right)\left[\left(r^{2}+a^{2}\right)E-al_{z}\right]}{\Delta}+a\left(l_{z}-aE\sin^{2}{\theta}\right)+\epsilon_{\mu}\epsilon_{r}\dfrac{2Mr}{\Delta}\sqrt{R(r)}, \\
    \dfrac{d\tilde{\phi}}{d\tilde{s}}&=\dfrac{a\left[\left(r^{2}+a^{2}\right)E-al_{z}\right]}{\Delta}+\dfrac{1}{\sin^{2}{\theta}}\left(l_{z}-aE\sin^{2}{\theta}\right)+\epsilon_{\mu}\epsilon_{r}\dfrac{a}{\Delta}\sqrt{R(r)},
\end{align}
\end{subequations}
while the radial and polar equations remain unaffected.

For timelike geodesics, it is convenient to use dimensionless variables $\tilde{\tau}$, $\xi$, $\alpha$, $\varepsilon$, $\lambda_z$, $\kappa$, and $s$, defined as
\begin{align}
\label{scaling}
\tilde{\tau}=\dfrac{\tilde{t}}{M}, \qquad \xi=\dfrac{r}{M}, \qquad \alpha=\dfrac{a}{M}, \qquad \varepsilon=\dfrac{E}{m}, \qquad \lambda_{z}=\dfrac{l_{z}}{Mm}, \qquad \kappa=\dfrac{\mathcal{K}}{M^{2}m^{2}}, \qquad s=Mm\tilde{s}.
\end{align}
This gives
\begin{align}
    R(r) = M^4 m^2 \tilde R(\xi), \qquad \Theta(\theta) = M^2 m^2 \tilde \Theta(\theta),
\end{align}
where
\begin{subequations}
\label{potentials}
\begin{align}
    \tilde{R}(\xi)&=\left[\left(\xi^{2}+\alpha^{2}\right)\varepsilon-\alpha\lambda_{z}\right]^{2}-\left(\xi^{2}-2\xi+\alpha^{2}\right)\left(\kappa+\xi^{2}\delta\right), \\
    \tilde{\Theta}(\theta)&=\kappa-\alpha^{2}\delta\cos^{2}{\theta}-\left(\dfrac{\lambda_{z}}{\sin{\theta}}-\alpha\varepsilon\sin{\theta}\right)^{2}.
\end{align}
\end{subequations}
For null geodesics, for which $m = 0$, the mass parameter $m$ in Eqs.\ \eqref{scaling}--\eqref{potentials} can be replaced with any parameter $\tilde m > 0$. In terms of dimensionless quantities, geodesic equations \eqref{geoeq2} in horizon-penetrating coordinates can be written as
\begin{subequations}
\begin{align}
    \dfrac{d\tilde{\tau}}{ds}&=\dfrac{\left(\xi^{2}+\alpha^{2}\right)\left[\left(\xi^{2}+\alpha^{2}\right)\varepsilon-\alpha\lambda_{z}\right]}{\xi^{2}-2\xi+\alpha^{2}}+\alpha\left(\lambda_{z}-\alpha\varepsilon\sin^{2}{\theta}\right)+\epsilon_{\mu}\epsilon_{r}\dfrac{2\xi}{\xi^{2}-2\xi+\alpha^{2}}\sqrt{\tilde{R}(\xi)}, \label{timedimless} \\
    \dfrac{d\xi}{ds}&=\epsilon_{r}\sqrt{\tilde{R}(\xi)}, \label{radialdimless} \\
    \dfrac{d\theta}{ds}&=\epsilon_{\theta}\sqrt{\tilde{\Theta}(\theta)}, \label{polardimless} \\
    \dfrac{d\tilde{\phi}}{ds}&=\dfrac{\alpha\left[\left(\xi^{2}+\alpha^{2}\right)\varepsilon-\alpha\lambda_{z}\right]}{\xi^{2}-2\xi+\alpha^{2}}+\dfrac{1}{\sin^{2}{\theta}}\left(\lambda_{z}-\alpha\varepsilon\sin^{2}{\theta}\right)+\epsilon_{\mu}\epsilon_{r}\dfrac{\alpha}{\xi^{2}-2\xi+\alpha^{2}}\sqrt{\tilde{R}(\xi)}. \label{phidimless}
\end{align}
\end{subequations}
For $\epsilon_\mu = +1$ (ingoing coordinates) these equations were derived (and solved) in \cite{B2025}. In what follows, we briefly discuss our solving method, mainly because we now cover both ingoing and outgoing coordinates.

\section{Solutions of geodesic equations}
\label{sec:geosol}
\subsection{Biermann-Weierstrass formula}

The radial and polar equations \eqref{radialdimless} and \eqref{polardimless} can be solved straightforwardly, using an old but largely forgotten result by Biermann and Weierstrass \cite{BW1865}. It was used in \cite{CM2022} to describe causal geodesics in the Schwarzschild spacetime and subsequently applied to Kerr geodesics in Boyer-Lindquist coordinates in \cite{CMH2023} and in ingoing horizon-penetrating coordinates in \cite{B2025}. Since in this work we deal with geodesics plunging into the black hole, as well as those emanating from the ``white hole'', we add to the description of \cite{B2025} formulas valid for outgoing horizon-penetrating coordinates.

The Biermann-Weierstrass theorem can be stated as follows. Let $f(x)$ be a quartic polynomial
\begin{align}
f(x)=a_{0}x^{4}+4a_{1}x^{3}+6a_{2}x^{2}+4a_{3}x+a_{4}
\end{align}
and let $g_2$ and $g_3$ denote the associated Weierstrass invariants
\begin{subequations}
\begin{align}
g_{2}&=a_{0}a_{4}-4a_{1}a_{3}+3a_{2}^{2},\\    g_{3}&=a_{0}a_{2}a_{4}+2a_{1}a_{2}a_{3}-a_{2}^{3}-a_{0}a_{3}^{2}-a_{1}^{2}a_{4}.
\end{align}
\end{subequations}
Define
\begin{align}    z=\int\limits_{x_{0}}^{x}\dfrac{d\bar{x}}{\sqrt{f(\bar{x})}},
\end{align}
where $x_0$ denotes any constant. Then, the relation $z = z(x)$ can be inverted. The value $x = x(z)$ is given by
\begin{align}
    x=x_{0}+\dfrac{-\sqrt{f(x_{0})}\wp'(z)+\frac{1}{2}f'(x_{0})\left[\wp(z)-\frac{1}{24}f''(x_{0})\right]+\frac{1}{24}f(x_{0})f'''(x_{0})}{2\left[\wp(z)-\frac{1}{24}f''(x_{0})\right]^{2}-\frac{1}{48}f(x_{0})f^{\left(4\right)}(x_{0})},
\end{align}
where the Weierstrass function $\wp$ is computed for the invariants $g_2$ and $g_3$, i.e., $\wp(z)=\wp(z;g_{2},g_{3})$, and where the prime denotes differentiation. The proofs can be found in \cite{BW1865,Greenhill1892,Reynolds1989,CM2022}.

\subsection{Radial solution}

The dimensionless radial effective potential $\tilde{R}(\xi)$ can be expressed as
\begin{align}
    \tilde{R}(\xi)=a_{0}\xi^{4}+4a_{1}\xi^{3}+6a_{2}\xi^{2}+4a_{3}\xi+a_{4},
\end{align}
where
\begin{subequations}
\begin{align}
    a_{0}&=\hphantom{+}\varepsilon^{2}-\delta, \\ \rule{0pt}{1.66em}
    a_{1}&=\hphantom{+}\dfrac{1}{2}\delta, \\ \rule{0pt}{1.66em}
    a_{2}&=-\dfrac{1}{6}\left(\alpha^{2}\delta+\kappa-2\alpha^{2}\varepsilon^{2}+2\alpha\varepsilon\lambda_{z}\right), \\ \rule{0pt}{1.66em}
    a_{3}&=\hphantom{+}\dfrac{1}{2}\kappa, \\ \rule{0pt}{1.66em}
    a_{4}&=-\alpha^{2}\left[\kappa-\left(\alpha\varepsilon-\lambda_{z}\right)^{2}\right].
\end{align}
\end{subequations}
The above coefficients determine the Weierstrass invariants as follows:
\begin{subequations}
\begin{align}
    g_{\tilde{R},2}&=a_{0}a_{4}-4a_{1}a_{3}+3a_{2}^{2}, \\
    g_{\tilde{R},3}&=a_{0}a_{2}a_{4}+2a_{1}a_{2}a_{3}-a_{2}^{3}-a_{0}a_{3}^{2}-a_{1}^{2}a_{4}.
\end{align}
\label{WP_g_inv}
\end{subequations}
The application of the Biermann-Weierstrass theorem to Eq.\ \eqref{radialdimless} yields the solution in the form
\begin{align}
\label{xisol}
    \xi(s)=\xi_{0}+\dfrac{-\epsilon_{r,0}\sqrt{\tilde{R}(\xi_{0})}\wp_{\tilde{R}}'(s)+\frac{1}{2}\tilde{R}'(\xi_{0})\left[\wp_{\tilde{R}}(s)-\frac{1}{24}\tilde{R}''(\xi_{0})\right]+\frac{1}{24}\tilde{R}(\xi_{0})\tilde{R}'''(\xi_{0})}{2\left[\wp_{\tilde{R}}(s)-\frac{1}{24}\tilde{R}''(\xi_{0})\right]^{2}-\frac{1}{48}\tilde{R}(\xi_{0})\tilde{R}^{\left(4\right)}(\xi_{0})},
\end{align}
where $\wp_{\tilde R}(s)=\wp(s;g_{\tilde{R},2},g_{\tilde{R},3})$ is the Weierstrass function defined by invariants \eqref{WP_g_inv} and $\xi_{0}$ denotes the initial radial position of the particle corresponding to $s=0$. The sign $\epsilon_{r,0} = \pm 1$ appearing in Eq.\ \eqref{xisol} corresponds to the radial direction of motion at $s = 0$, i.e., it is part of the initial data. 

\subsection{Polar solution}

The polar equation \eqref{polardimless} can be solved in a similar manner. The substitution $\mu = \cos \theta$ yields
\begin{align}
\dfrac{d\mu}{ds}=-\epsilon_{\theta}\sqrt{\hat{\Theta}(\mu)},
\end{align}
where the dimensionless polar effective potential reads
\begin{align}
    \begin{aligned}
        \hat{\Theta}(\mu)&=\left(1-\mu^{2}\right)\left(\kappa-\alpha^{2}\mu^{2}\delta-\dfrac{\lambda_{z}}{1-\mu^{2}}+2\alpha\varepsilon\lambda_{z}-\alpha^{2}\varepsilon^{2}+\alpha^{2}\varepsilon^{2}\mu^{2}\right) \\
        &=b_{0}\mu^{4}+6b_{2}\mu^{2}+b_{4},
    \end{aligned}
\end{align}
with
\begin{subequations}
\begin{align}
    b_{0}&=-\alpha^{2}a_{0}, \\ \rule{0pt}{1.5em}
    b_{2}&=\hphantom{+}a_{2}, \\ \rule{0pt}{1.5em}
    b_{4}&=-\dfrac{1}{\alpha^{2}}a_{4}.
\end{align}
\end{subequations}
The Weierstrass invariants corresponding to $\hat \Theta(\mu)$ (a quartic polynomial in $\mu$) are expressed as
\begin{subequations}  
\begin{align}
g_{\hat{\Theta},2}&=b_{0}b_{4}+3b_{2}^{2}, \\
g_{\hat{\Theta},3}&=b_{0}b_{2}b_{4}-b_{2}^{3}.
\end{align}
\end{subequations}
The Biermann-Weierstrass formula yields the solution for $\mu(s)$ in the form
\begin{align}
\mu(s)=\mu_{0}+\dfrac{\epsilon_{\theta,0}\sqrt{\hat{\Theta}(\mu_{0})}\wp_{\hat{\Theta}}'(s)+\frac{1}{2}\hat{\Theta}'(\mu_{0})\left[\wp_{\hat{\Theta}}(s)-\frac{1}{24}\hat{\Theta}''(\mu_{0})\right]+\frac{1}{24}\hat{\Theta}(\mu_{0})\hat{\Theta}'''(\mu_{0})}{2\left[\wp_{\hat{\Theta}}(s)-\frac{1}{24}\hat{\Theta}''(\mu_{0})\right]^{2}-\frac{1}{48}\hat{\Theta}(\mu_{0})\hat{\Theta}^{\left(4\right)}(\mu_{0})},
\end{align}
where $\wp_{\hat{\Theta}}(s)=\wp(s;g_{\hat{\Theta},2},g_{\hat{\Theta},3})$. Here $\mu_0 = \mu(0)$ and $\mu_0 = \cos \theta_0$, where $\theta_0 = \theta(0)$. Similarly to $\epsilon_{r,0}$ appearing in Eq.~\eqref{xisol}, the sign $\epsilon_{\theta,0} = \pm 1$ is fixed and corresponds to the polar direction of motion at $s = 0$. 

\subsection{Azimuthal solution}

Given solutions to the radial and polar equations, Eqs.\ \eqref{timedimless} and \eqref{phidimless} can be integrated by quadratures. In principle, this can be done exactly, and the relevant formulas can be found in \cite{CMH2023}. Since the resulting formulas are quite lengthy, we perform the integrals for $\tilde \phi$ and $\tilde \tau$ numerically.

Note that the integrands in the expression
\begin{align}
    \begin{aligned}        \tilde{\phi}(s)&=\tilde{\phi}_{0}+\int\limits^{s}_0\left[\dfrac{\alpha\left[\left(\xi^{2}(\bar{s})+\alpha^{2}\right)\varepsilon-\alpha\lambda_{z}\right]}{\xi^{2}(\bar{s})-2\xi(\bar{s})+\alpha^{2}}+\dfrac{\lambda_{z}-\alpha\varepsilon\sin^{2}{\theta(\bar{s})}}{\sin^{2}{\theta(\bar{s})}}\right]d\bar{s} \\
    &\hphantom{=}+\int\limits^{s}_0\dfrac{\epsilon_{\mu}\epsilon_{r}\alpha}{\xi^{2}(\bar{s})-2\xi(\bar{s})+\alpha^{2}}\sqrt{\tilde{R}\left(\xi(\bar{s})\right)}d\bar{s}
    \end{aligned}
    \label{almost_phi_sol}
\end{align}
diverge at the horizons $\xi_{\pm}=1\pm\sqrt{1-\alpha^{2}}$. This divergent behavior can be removed, depending on the signs $\epsilon_\mu$ and $\epsilon_r$. For the incoming coordinates ($\epsilon_\mu = +1$), this was described in \cite{B2025}. We first write \eqref{almost_phi_sol} in the form
\begin{align}
\tilde{\phi}(s)=\tilde{\phi}_{0}+\lambda_{z}\int\limits^{s}_0\dfrac{d\bar{s}}{\sin^{2}{\theta(\bar{s})}}+\alpha\int\limits^{s}_0\dfrac{2\varepsilon\xi(\bar{s})-\alpha\lambda_{z}+\epsilon_{\mu}\epsilon_{r}\sqrt{\tilde{R}\left(\xi(\bar{s})\right)}}{\xi^{2}(\bar{s})-2\xi(\bar{s})+\alpha^{2}}d\bar{s}.
\end{align}
Noting that
\begin{align}
\dfrac{\left(\xi^{2}+\alpha^{2}\right)\varepsilon-\alpha\lambda_{z}+\epsilon_{\mu}d\xi/ds}{\xi^{2}-2\xi+\alpha^{2}}=\dfrac{\delta\xi^{2}+\kappa}{\left(\xi^{2}+\alpha^{2}\right)\varepsilon-\alpha\lambda_{z}-\epsilon_{\mu}d\xi/ds},
\end{align}
we obtain
\begin{align}
\tilde{\phi}(s)=\tilde{\phi}_{0}+\lambda_{z}\int\limits^{s}_0\dfrac{d\bar{s}}{\sin^{2}{\left(\theta(\bar{s})\right)}}+\alpha\int\limits^{s}_0\left[\dfrac{\delta\xi^{2}(\bar{s})+\kappa}{\left(\xi^{2}(\bar{s})+\alpha^{2}\right)\varepsilon-\alpha\lambda_{z}-\epsilon_{\mu} d\xi(\bar{s})/d\bar{s}}-\varepsilon\right]d\bar{s}.
\end{align}
If $\left(\xi^{2}+\alpha^{2}\right)\varepsilon-\alpha\lambda_{z}-\epsilon_{\mu} d\xi/ds \neq 0$, the above expression remains explicitly regular at the horizons. This, in turn, depends on the signs $\epsilon_\mu$ and $\epsilon_r$.

Note that the condition $\left(\xi^{2}+\alpha^{2}\right)\varepsilon-\alpha\lambda_{z}-\epsilon_{\mu} d\xi/ds = 0$ can also be written as
\begin{equation}
\label{eqdiverge}
A = \epsilon_\mu \epsilon_r \sqrt{\tilde R(\xi)},
\end{equation}
where $A \coloneqq \left(\xi^{2}+\alpha^{2}\right)\varepsilon-\alpha\lambda_{z}$. By computing the squares of both sides and subtracting the resulting expressions, we see that this equation implies that
\begin{equation}
\label{squared}
    A^2 - \tilde R(\xi) = \left( \xi^2 - 2 \xi + \alpha^2 \right) \left( \delta \xi^2 + \kappa \right) = 0.
\end{equation}
Consequently, Eq.\ \eqref{eqdiverge} can only be satisfied at the horizons $\xi = \xi_\pm$ (we only consider the case with $\delta \xi^2 + \kappa > 0$). Moreover, this can only happen if the sign of $A$ is equal to $\epsilon_\mu \epsilon_r$. We should also stress that the first equality in Eq.\ \eqref{squared} holds generally, independent of Eq.\ \eqref{eqdiverge}. Consequently, $A^2 - \tilde R(\xi)$ has to vanish at $\xi = \xi_\pm$. Since
\begin{equation}
    A^2 - \tilde R(\xi) = \left( A -  \epsilon_\mu \epsilon_r \sqrt{\tilde R(\xi)} \right) \left( A +  \epsilon_\mu \epsilon_r \sqrt{\tilde R(\xi)} \right),
\end{equation}
we conclude that at each of the horizons, out of two cases $\epsilon_\mu \epsilon_r = \pm 1$, one corresponds to a regular behavior and the second to a geodesic with divergent $\tilde \phi$.
 
The sign of $A$ depends on the time orientation of the geodesic. In Blocks I and II, in which the vector $X$ defined in Eq.\ \eqref{XY} is future-directed, we have
\begin{equation}
    g(p,X) = (r^2 + a^2) p_{\tilde t} + a p_{\tilde \phi} = - \left[ (r^2 + a^2) E - a l_z \right] = - M^2 m \left[ (\xi^2 + \alpha^2) \varepsilon - \alpha \lambda_z \right] = - M^2 m A.
\end{equation}
Thus, a future-directed timelike geodesic is characterized by $A > 0$ in Blocks I and II. In Block II, we check
\begin{equation}
    g(p,Y) =  2 M r p_{\tilde t} + \epsilon_\mu \Delta p_r + a p_{\tilde \phi} = - 2 M r E + \epsilon_\mu \Delta p_r + a l_z=\epsilon_{\mu}\epsilon_{r}\sqrt{R(r)},
\end{equation}
where in the last equality, we have used the fact that $\Delta p_r = \epsilon_r \sqrt{R(r)} + \epsilon_\mu \left( 2 M r E - a l_z \right)$. Consequently, a future-pointing geodesic must be characterized by $\epsilon_\mu \epsilon_r = -1$ in Block II. This result also shows that a timelike geodesic cannot have a radial turning point for $r \in (r_-,r_+)$. In particular, a future-pointing timelike geodesic plunging into the black hole from Block I (where $A > 0$), passes through Block II and enters Block III (where again $A > 0$). Since $A$ depends on the radius only through the term $\varepsilon \xi^2$, we conclude that, for such a trajectory, $A > 0$ also in Block II. That also means that for this trajectory, one has to choose $\epsilon_\mu \epsilon_r = -1$ to prevent the divergent behavior of $\tilde \phi$ at the horizons.

\subsection{Time solution}

A similar procedure can be applied to the integral expressions for the time coordinate
\begin{align}
\begin{aligned}
    \tilde{\tau}(s)&=\tilde{\tau}_{0}+\int\limits^{s}_{0}\left[\dfrac{\left(\xi^{2}(\bar{s})+\alpha^{2}\right)\left[\left(\xi^{2}(\bar{s})+\alpha^{2}\right)\varepsilon-\alpha\lambda_{z}\right]}{\xi^{2}(\bar{s})-2\xi(\bar{s})+\alpha^{2}}+\alpha\left(\lambda_{z}-\alpha\varepsilon\sin^{2}{\theta(\bar{s})}\right)\right]d\bar{s} \\
    &\hphantom{=}+\int\limits^{s}_{0}\dfrac{2\epsilon_{\mu}\epsilon_{r}\xi(\bar{s})}{\xi^{2}(\bar{s})-2\xi(\bar{s})+\alpha^{2}}\sqrt{\tilde{R}\left(\xi(\bar{s})\right)}d\bar{s},
\end{aligned}    
\end{align}
which can be expressed as
\begin{align}
    \tilde{\tau}(s)=\tilde{\tau}_{0}-\alpha^{2}\varepsilon\int\limits^{s}_{0}\sin^{2}{\theta(\bar{s})}d\bar{s}+\int\limits^{s}_{0}\dfrac{\left(\xi^{2}(\bar{s})+\alpha^{2}\right)^{2}\varepsilon-2\alpha\lambda_{z}\xi(\bar{s})+2\xi(\bar{s})\epsilon_{\mu}\epsilon_{r}\sqrt{\tilde{R}\left(\xi(\bar{s})\right)}}{\xi^{2}(\bar{s})-2\xi(\bar{s})+\alpha^{2}}d\bar{s}.
\end{align}
We now note that
\begin{align}
\dfrac{\left(\xi^{2}+\alpha^{2}\right)^{2}\varepsilon-2\alpha\lambda_{z}+2\xi\epsilon_{\mu}d\xi/ds}{\xi^{2}-2\xi+\alpha^{2}}=\varepsilon\left(\xi^{2}+\alpha^{2}\right)+\dfrac{2\xi\left(\delta\xi^{2}+\kappa\right)}{\left(\xi^{2}+\alpha^{2}\right)\varepsilon-\alpha\lambda_{z}-\epsilon_{\mu}d\xi/ds}.
\end{align}
This leads to the following representation of the solution $\tilde{\tau}(s)$, regular at the horizons $\xi_{\pm}$:
\begin{align}
\label{tauregular}
    \tilde{\tau}(s)=\tilde{\tau}_{0}-\alpha^{2}\varepsilon\int\limits^{s}_{0}\sin^{2}{\left(\theta(\bar{s})\right)}d\bar{s}+\int\limits^{s}_{0}\left[\left(\xi^{2}(\bar{s})+\alpha^{2}\right)\varepsilon+\dfrac{2\xi(\bar{s})\left(\delta\xi^{2}(\bar{s})+\kappa\right)}{\left(\xi^{2}(\bar{s})+\alpha^{2}\right)\varepsilon-\alpha\lambda_{z}-\epsilon_{\mu} d\xi(\bar{s})/d\bar{s}}\right]d\bar{s}.
\end{align}

We emphasize that the potentially problematic denominator in Eq.\ \eqref{tauregular} has the form $A - \epsilon_\mu d\xi/ds$, which has already been discussed in the context of the coordinate $\tilde \phi$. The choice of $\epsilon_\mu \epsilon_r$ leading to the regular behavior of $\tilde \tau$ also leads to the regular $\tilde \phi$.

\section{Causal geodesics on COS diagrams}
\label{sec:examples}
In this section we provide explicit examples in which timelike Kerr geodesics are projected onto COS diagrams. Geodesic equations were solved using \textit{Wolfram Mathematica} \cite{Wolfram} to generate pairs of points $(\tilde{\tau},\xi)$, which were subsequently mapped onto the COS projection diagram using the transformations described in Sec.\ \ref{sec:projdiag} and combined with the horizon-penetrating coordinates employed in Sec.\ \ref{sec:geoeq}. To this end, the transformations between the $(t,r)$ and $(T,X)$ coordinates were modified by substituting the horizon-penetrating time, defined according to Eq.\ \eqref{ttilde}, and subsequently combining it with the dimensionless coordinates $\tilde{\tau}$ and $\xi$.

\begin{figure}[t]
\centering
  \begin{minipage}[c]{0.35\textwidth}
    \centering {\includegraphics[width=\textwidth]{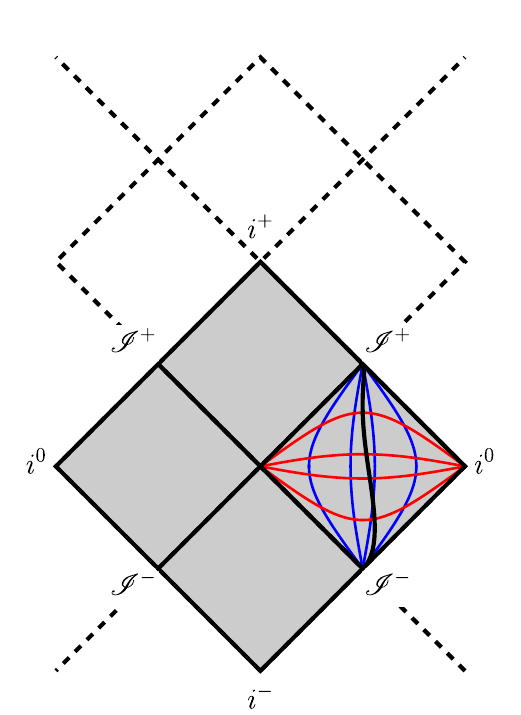}}
  \end{minipage}\hspace{1cm}
  \begin{minipage}[c]{0.45\textwidth}
    \centering{\includegraphics[width=1\textwidth]{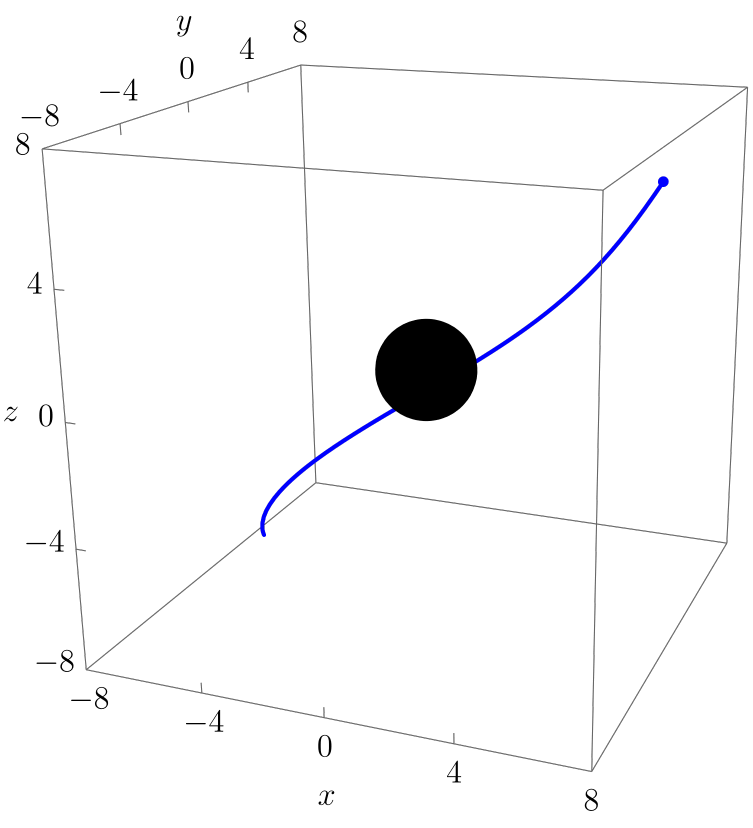}}
  \end{minipage}
\caption{\label{fig:example1} A timelike geodesic on the COS projection diagram for the Kerr spacetime and its three-dimensional $\mathbb{R}^{3}$ counterpart. The red lines on the diagram denote hypersurfaces of constant Boyer-Lindquist time $t=\left\{-5,-1,1,5\right\}$, whereas the blue lines represent hypersurfaces of constant radial coordinate corresponding to $f(r)=\left\{-5,-1,1,5\right\}$.}
\end{figure}

The first example illustrates a trajectory entirely contained within Region 1 (and within Block I on the COS diagram). Consequently, it can be described using a single (ingoing) coordinate system with $\epsilon_\mu = +1$. The parameters of this geodesic are listed in Tab.~\ref{tab:parameters1}. We set $\alpha = 0.8$. The Mino time spans the range $s \in (0,s_\mathrm{max})$, where $s_{\mathrm{max}}=1$. The resulting geodesic is illustrated in Fig.\ \ref{fig:example1}. The left panel shows its projection onto the COS diagram, as described in previous sections. Since the geodesic in this example does not cross any of the horizons, there is no direct need to achieve smooth gluing between different blocks of the diagram. We apply the rules outlined in Sec.\ \ref{sec:projdiag} mostly for illustration purposes. The relevant parameters are chosen as follows. The constant $k$ is chosen as $k = k_+$, allowing for a smooth gluing along $r = r_+$. For $\alpha=0.8$ one has $k_{+}=0.1875$. We choose $r_{1}=5/2$. The constant $f_{0,1}$ is fixed by requiring that $f(\xi(s_{\mathrm{max}}/2))=0$. This yields $f_{0,1}=4.76132$. The four Blocks I, II, I$'$, and II$'$ that can be smoothly glued together are colored  gray.

The right panel in Fig.\ \ref{fig:example1} depicts a projection of the same geodesic onto a hypersurface $\tilde t = \mathrm{const}$. This plot uses Cartesian coordinates defined as
\begin{align}
    \begin{aligned}
        \left\{\begin{array}{rl}
            x&=\xi\sin{\theta}\cos{\phi}, \\
            y&=\xi\sin{\theta}\sin{\phi}, \\
            z&=\xi\cos{\theta}.
        \end{array}\right.
    \end{aligned}
\end{align}
Note that they differ from standard Kerr-Schild Cartesian coordinates. The starting point $(x_{0},y_{0},z_{0})$, corresponding to $(\xi_{0},\theta_{0},\phi_{0})$, is marked with a blue dot. The surface of the outer horizon of the Kerr black hole, defined by $\xi=\xi_{+}=1.6$, is depicted in black. 

\begin{table}[t]
\begin{ruledtabular}
\caption{\label{tab:parameters1} Parameters of the timelike geodesic illustrated in Fig.\ \ref{fig:example1}.}
    \begin{tabular}{lccccccccccc}
        Type of coordinates & $\hphantom{+}\tilde{\tau}_{0}$ & $\xi_{0}$ & $\theta_{0}$ & $\tilde{\phi}_{0}$ & $\epsilon_{\mu}$ & $\epsilon_{r,0}$ & $\epsilon_{\theta,0}$ & $\alpha$ & $\varepsilon$ & $\lambda_{z}$ & $\kappa$ \\
        \colrule
        Ingoing coordinates & $-20$ & $10$ & $0.85$ & $0$ & $+1$ & $-1$ & $+1$ & $0.8$ & $\sqrt{1.1}$ & $3.0$ & $12$
    \end{tabular}
\end{ruledtabular}
\end{table}

As the next example, we chose probably the most challenging case of a geodesic that starts in Block I, plunges into the black hole, traverses through Block II, and encounters a radial turning point in Block III, beneath the inner horizon. In the extended Kerr spacetime, a test particle moving along such a geodesic can then escape through the Cauchy horizon and the horizon of the white hole. The parameters chosen for this example are collected in Table~\ref{tab:parameters}. We set $\alpha = 0.9$. Consequently, the horizons are located at $\xi_{+}=1.43589$ and $\xi_{-}=0.56411$. The segment of the trajectory plunging into the black hole is described in ingoing coordinates. The particle starting at $\xi_0 = 10$, crosses the horizons located at $\xi_+$ and $\xi_-$ at Mino times $s_{+}^{\mathrm{in}}=0.906616$ and $s_{-}^{\mathrm{in}}=1.52113$, respectively. The radial turning point at $\xi_{\mathrm{turn}}=0.552326$ is reached in Mino time $s_{\mathrm{turn}}=1.59333$. At this point, we switch to outgoing coordinates, with $\epsilon_\mu = -1$. We emphasize that the choice of the location at which one switches from ingoing to outgoing coordinates is to a large extent arbitrary, but this transition must be made in Block III, as the ingoing coordinates diverge for the outgoing trajectory approaching $\xi = \xi_-$. The outgoing geodesic traverses the horizons located at $\xi_{-}$ and $\xi_{+}$ in reverse order at Mino times $s_{-}^{\mathrm{out}}=1.66553$ and $s_{+}^{\mathrm{out}}=2.28005$. It should be noted that the initial coordinates of the test particle for the outgoing case correspond to the values of the respective coordinates at $s_{\mathrm{turn}}$, with the exception of time $\tilde{\tau}$. The latter was adjusted to ensure a continuous gluing between the ingoing and outgoing geodesic segments on the diagram. This can be achieved by setting $\tilde{\tau}$ so that the Boyer-Lindquist time $t=\tilde{t}-\epsilon_{\mu}(r_{\ast}-r) -C_{t}$ remains consistent before and after reaching the turning point. The geodesic shown in the diagram was terminated at $s_{\mathrm{max}}=3.33$.

\begin{table}[t]
\begin{ruledtabular}
\caption{\label{tab:parameters} Parameters of the timelike geodesic illustrated in Fig.\ \ref{fig:3-geo}.}
    \begin{tabular}{lccccccccccc}
        Type of coordinates & $\hphantom{+}\tilde{\tau}_{0}$ & $\xi_{0}$ & $\theta_{0}$ & $\tilde{\phi}_{0}$ & $\epsilon_{\mu}$ & $\epsilon_{r,0}$ & $\epsilon_{\theta,0}$ & $\alpha$ & $\varepsilon$ & $\lambda_{z}$ & $\kappa$ \\
        \colrule
        Ingoing coordinates & $-20$ & $10$ & $0.85$ & $0$ & $+1$ & $-1$ & \multirow{2}{*}{+1} & \multirow{2}{*}{0.90} & \multirow{2}{*}{$\sqrt{1.18}$} & \multirow{2}{*}{$1.00$} & \multirow{2}{*}{$9.00$} \\
        \cline{1-7}
        Outgoing coordinates & $-3.81932$ & $0.552326$ & $0.531808$ & $2\pi+1.89936$ & $-1$ & $+1$ &  & &  \\
    \end{tabular}
\end{ruledtabular}
\end{table}

\begin{figure}[H]
\centering
{\captionsetup[subfloat]{labelformat=empty}
  \begin{minipage}[c]{0.35\textwidth}
    \centering
    \subfloat[b)]{\includegraphics[width=\textwidth]{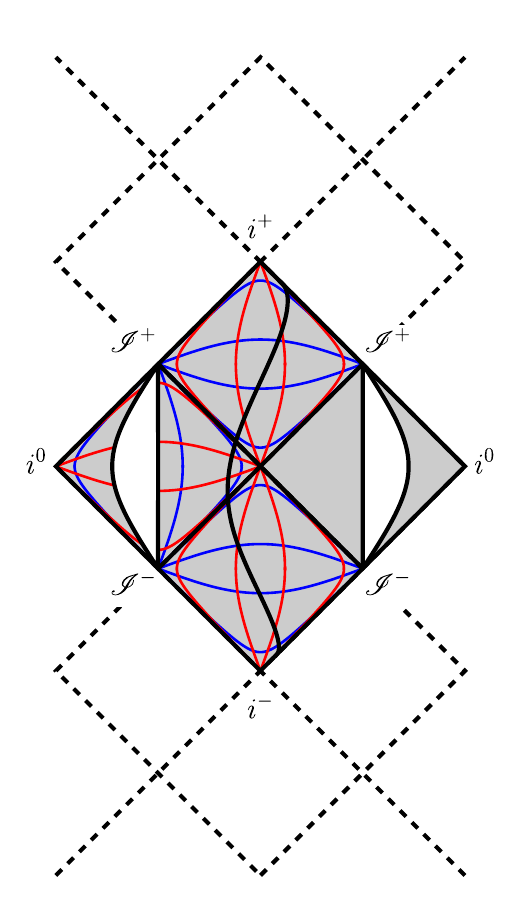}}
  \end{minipage}
  \begin{minipage}[c]{0.35\textwidth}
    \centering
    \subfloat[c)]{\includegraphics[width=\textwidth]{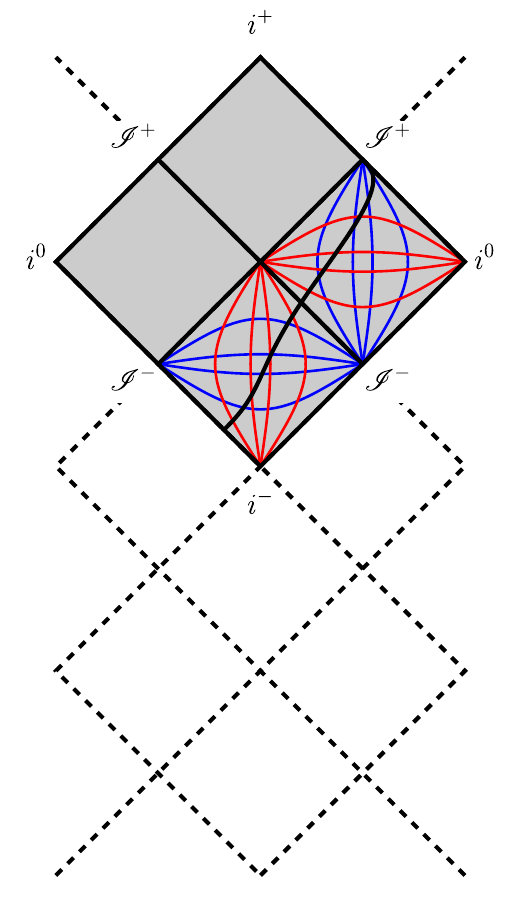}}\\
    \subfloat[a)]
    {\includegraphics[width=\textwidth]{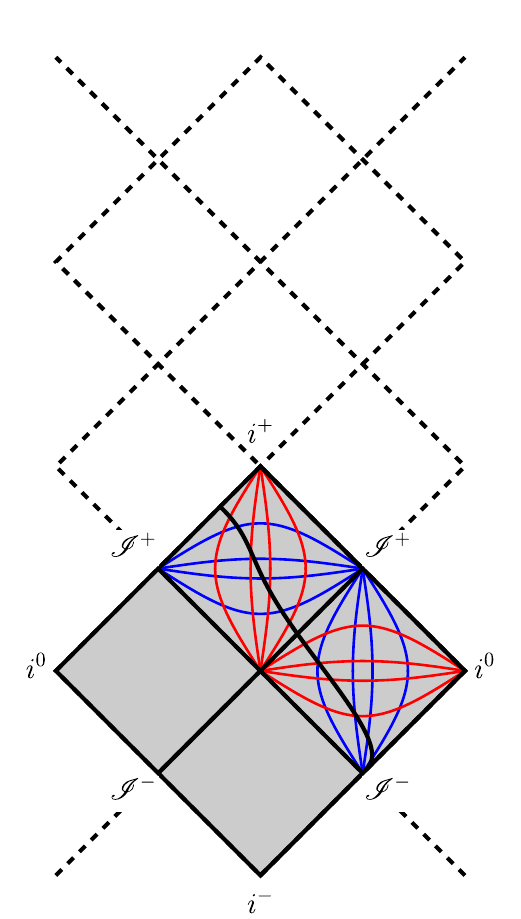}}
  \end{minipage}
}
\caption{\label{fig:3-geo} A timelike geodesic on the COS projection diagram for the Kerr spacetime. The red lines depict hypersurfaces of constant Boyer-Lindquist time $t=\left\{-5,-1,1,5\right\}$, whereas the blue ones represent hypersurfaces of constant radial coordinate $r$ corresponding to $f(r)=\left\{-5,-1,1,5\right\}$.}
\end{figure}

The diagram shown in Fig.~\ref{fig:3-geo}, which illustrates this geodesic, is constructed for $M=1$ and $a=0.9$. To ensure the continuity of projected geodesics across the horizons, we have adopted constants $r_{i}$ and $f_{0,i}$ according to Table~\ref{tab:constants}. As discussed in Sec.\ \ref{sec:projdiag}, the constants $f_{0,4}$ in Region 4 ($-\infty < r <\hat{r}$) and $f_{0,3}$ in Region 3 ($0 < r < r_-$) can be chosen independently of each other, even though both Regions 3 and 4 belong to the same Block III in the diagram. On the other hand, the constant $f_{0,3}$ must be set so that $f(\hat{r})\leq f(0)$, to ensure that the hypersurfaces of constant $r$ are plotted in the correct order.

\begin{table}[t]
\begin{ruledtabular}
\caption{\label{tab:constants} Numerical parameters used in constructing the COS projection diagrams shown in Fig.\ \ref{fig:3-geo}.}
\label{tab:const}
    \begin{tabular}{cccccc}
        $k$ & Region & $r_{i}$ & $\hphantom{+}f_{0,i}$ & $\hphantom{+}\hat{f}(r_{\pm})$ & \\
        \colrule
        \multirow{2}{*}{$k_{+}=\hphantom{+}0.151784$} & $(r_{+},\infty)$ & $2.50$ & $\hphantom{+}0.00$ & \multirow{2}{*}{$\hphantom{+}1.05241$} & \\
        \cline{2-4}
        & $(r_{-},r_{+})$ & $0.66$ & $\hphantom{+}1.16314$ & & \\
        \colrule
        \multirow{3}{*}{$k_{-}=-0.386352$} & $(r_{-},r_{+})$ & $0.66$ & $\hphantom{+}0.483149$ & \multirow{3}{*}{$-1.37869$} & \\
        \cline{2-4}
        & $(0,r_{-})$ & $0.33$ & $\hphantom{+}0.415833$ & & \\
        \cline{2-4}
        & $(-\infty,\hat{r})$ & $\hat{r}-0.33$ & $-2.063049$ & &
    \end{tabular}
\end{ruledtabular}
\end{table}

Since only four conformal blocks can be glued smoothly, the depiction of the entire geodesic requires three separate diagrams, labeled in Fig.\ \ref{fig:3-geo}, as a), b), and c). They are constructed using both conformal factors $\Phi^2$ (diagrams a) and c)) and $\Psi^2$ (diagram b)). The white region in Blocks III and III$'$ in diagram b) coincides with the set $\mathcal V'$, with the boundaries corresponding to $r=\hat{r}=-0.948042$ and $r=0$. The constants $f_{0,3}$ in Blocks III and III$'$ were chosen in such a way that the locus $r=0$ forms a straight vertical line. The diagrams shown in Fig.\ \ref{fig:3-geo} should be read in the order of a), b), and c).

\section{Conclusions}
\label{sec:conclusions}

A representation of the geodesic motion on causal-structure-preserving spacetime diagrams is usually done in a schematic way (see \cite{Dyson} for an elegant example). This is mainly due to the fact that standard Carter-Penrose conformal diagrams are constructed for two-dimensional subspaces of the spacetime manifold. While this approach might be sufficient for highly-symmetric spacetimes, in particular spherically symmetric ones, it becomes problematic for less symmetric spacetime metrics, especially if the motion is not restricted to the two-dimensional manifold of the diagram. In principle, COS diagrams remove these obstacles, allowing for a representation of any causal curves. On the other hand, usual limitations related to standard gluing of individual causal blocks, used for Carter-Penrose diagrams, persist in the COS construction.

In this paper, we have shown examples of timelike geodesics in the extended Kerr spacetime, presented rigorously (in the numerical sense) on COS diagrams. While geodesics restricted to the black hole exterior can easily be presented on a single diagram, an exact depiction of a geodesic which passes through both Kerr horizons,  encounters a radial turning point beneath the Cauchy horizon and escapes from the white hole in subsequent blocks, may require additional diagrams (with distinct conformal factors). In this case, a numerical solution of geodesic equations requires two coordinate systems (ingoing and outgoing). While such examples lead to a complicated picture, we believe they serve as best illustrations of strengths and weaknesses of the COS method.

In all cases, COS diagrams can represent the full four-dimensional Kerr spacetime, apart from the region that must be excluded from the projection domain, containing the CTM set, i.e., they are not restricted to subspaces such as the axis ($\theta=0$) or the equatorial plane ($\theta=\pi/2$). A limitation of the COS construction for the Kerr spacetime is the presence of the CTM region in Block III, in which closed timelike curves exist, and which has to be removed from the COS diagram. Consequently, it is impossible to illustrate a timelike geodesic within the CTM set. In this respect, one should note that timelike Kerr geodesics can exhibit nontrivial behavior within the CTM region, where, in particular, radial turning points are allowed. We should also stress that the existence of the CTM region prevents a clear illustration of the difference between timelike geodesics terminating in the ring singularity (they are confined to the equatorial plane \cite{Neill1995}) and those that continue smoothly to $r < 0$. In both cases, the geodesic can only be illustrated up to $r = 0$.

It must also be emphasized that the exact shape of the images of a curve on the COS diagram is not uniquely defined. In particular, it depends on the choice of coordinates and is sensitive to integration constants such as $f_{0,i}$ and $r_{i}$, as discussed in Sec.\ \ref{sec:projdiag}. In the case of the Kerr spacetime, this drawback may act in conjunction with the existence of the CTM set. This happens, for instance, for the so-called transit Kerr geodesics, with no radial turning points. In the extended Kerr spacetime, they originate from $r = +\infty$, plunge into the black hole, and continue to $r = -\infty$, crossing Block III. In the COS diagram, the segment of such a geodesic within Block III is divided into two regions: Region 3 ($0 < r < r_-$) and Region 4 ($-\infty < r < \hat r$), disconnected by the CTM set. Since there is no unique way to connect the definitions of $r_{3}$, $f_{0,3}$ and $r_{4}$, $f_{0,4}$, corresponding to Regions 3 and 4, the images of transit geodesics in these regions remain unrelated to each other.

\begin{acknowledgments}
We would like to thank Sebastian Szybka for fruitful discussions.
\end{acknowledgments}

\end{document}